\documentclass[sigconf, nonacm, authorversion]{acmart}
\usepackage{graphicx}
\usepackage{microtype}
\usepackage{epstopdf}
\usepackage{flushend}
\usepackage{balance}
\usepackage{colortbl}
\usepackage{xcolor}
\usepackage{listings}
\usepackage{booktabs}
\usepackage{nicefrac}
\usepackage{paralist}
\usepackage{caption}
\usepackage{subcaption}
\usepackage{soul}
\usepackage{subfiles}
\usepackage{multirow}
\usepackage{enumitem}
\usepackage{tabularx}
\usepackage{xspace}
\usepackage{verbatim}
\usepackage{graphicx} 
\usepackage{pdfpages}

\newcommand{\fptree}{{FPTree}\xspace}
\newcommand{\lbtree}{{LB\textsuperscript{+}-Tree}\xspace}
\newcommand{\utree}{{$\mu$Tree}\xspace}
\newcommand{\dptree}{{DPTree}\xspace}
\newcommand{\pactree}{{PACTree}\xspace}
\newcommand{\hot}{{HOT}\xspace}
\newcommand{\masstree}{{Masstree}\xspace}
\newcommand{\roart}{{ROART}\xspace}
\newcommand{\nvtree}{{NV-Tree}\xspace}
\newcommand{\wbtree}{{wBTree}\xspace}
\newcommand{\pibench}{{PiBench}\xspace}
\newcommand{\bztree}{{BzTree}\xspace}

\newcommand{\jemalloc}{\texttt{jemalloc}\xspace}

\newcommand{\CAS}{\texttt{CAS}\xspace}

\newcommand{\CLWB}{{\tt CLWB}\xspace}
\newcommand{\CLFLUSHOPT}{{\tt CLFLUSHOPT}\xspace}

\newcommand{\load}{\texttt{load}\xspace}
\newcommand{\store}{\texttt{store}\xspace}

\def\thepaperkeywords{Persistent memory, PMem, indexing}
\def\thepapertitle{Evaluating Persistent Memory Range Indexes: Part Two [Extended Version]}

\hypersetup{
  pageanchor=true,
  plainpages=false,
  pdfpagelabels,
  bookmarks,
  bookmarksnumbered,
  pdfborder=0 0 0,  
  colorlinks=true,
  linkcolor=ACMRed,
  anchorcolor=ACMRed,
  citecolor=Blue, 
  pdftitle={\thepapertitle},
  pdfauthor={George He, Duo Lu, Kaisong Huang and Tianzheng Wang},
  pdfsubject={},
  pdfkeywords={\thepaperkeywords},
}

\usepackage{tikz}
\newcommand*\circled[1]{\tikz[baseline=(char.base)]{
            \node[shape=circle,fill=.,inner sep=0pt] (char) {\color{-.}\textsf\footnotesize #1};}}

\usepackage{breakurl}
\usepackage{algorithm}

\begin{document}

\renewcommand{\topfraction}{0.85}
\renewcommand{\textfraction}{0.1}
\renewcommand{\floatpagefraction}{0.80}


\definecolor{BrickRed}{rgb}{0.8, 0.25, 0.33}
\definecolor{comment-red}{rgb}{1,0,0}
\definecolor{OliveGreen}{rgb}{0,0.6,0}
\newcommand{\todo}[1]{\textnormal{\color{comment-red}{\textbf{TODO: #1}}}\unskip}
\newtheorem{mydef}{Invariant}
\newcommand{\algoname}[1]{{\textbf{#1}}}

\newcommand{\rdep}{\xrightarrow{\mathit{b}}}
\newcommand{\rdepstar}{\xrightarrow{\mathit{b*}}}
\newcommand{\fdep}{\xrightarrow{\mathit{f}}}
\newcommand{\wdep}{\xrightarrow{\mathit{w}}}
\newcommand{\xdep}{\rightarrow}
\newcommand{\txdepimpl}[3]{\xrightarrow{\mathit{#1{:}#2#3}}}
\newcommand{\txdep}[3][]{%
  \ifthenelse{\equal{#1}{}}
  {\txdepimpl{#2}{}{#3}}
  {\txdepimpl{#2}{#1{:}}{#3}}
}
\newcommand{\rwdep}[1][]{\txdep[#1]{r}{w}}
\newcommand{\wrdep}[1][]{\txdep[#1]{w}{r}}
\newcommand{\wwdep}[1][]{\txdep[#1]{w}{w}}
\newcommand{\wxdep}[1][]{\txdep[#1]{w}{x}}
\newcommand{\xxdep}[1][]{\txdep[#1]{x}{x}}


\title{\thepapertitle}
\thanks{This document is an extended version of ``Evaluating Persistent Memory Range Indexes: Part Two'' which will appear in Proceedings of the VLDB Endowment (PVLDB), Vol. 15, Issue 11 and will be presented at the 48th International Conference on
Very Large Data Bases (VLDB 2022). 
}

\lstset {
language=C,
basicstyle=\ttfamily\small,
keywordstyle=\ttfamily\bfseries\color{blue},
commentstyle=\color{OliveGreen},
numbers=left,
numberstyle=\small,
numbersep=5pt,
tabsize=1,
gobble=0,
stepnumber=2,
xleftmargin=15pt,
escapeinside={(@*}{*@)},
morekeywords={},
columns=fullflexible,
}


\settopmatter{authorsperrow=4}

\author{Yuliang He}
\affiliation{ \institution{Simon Fraser University} }
\email{georgeh@sfu.ca}

\author{Duo Lu}
\affiliation{ \institution{Simon Fraser University} }
\email{luduol@sfu.ca}

\author{Kaisong Huang}
\affiliation{ \institution{Simon Fraser University} }
\email{kha85@sfu.ca}

\author{Tianzheng Wang}
\affiliation{ \institution{Simon Fraser University} }
\email{tzwang@sfu.ca}

\begin{abstract}
Scalable persistent memory (PM) has opened up new opportunities for building indexes that operate and persist data directly on the memory bus, potentially enabling instant recovery, low latency and high throughput. 
When real PM hardware (Intel Optane Persistent Memory) first became available, previous work evaluated PM indexes proposed in the pre-Optane era. 
Since then, newer indexes based on real PM have appeared, but it is unclear how they compare to each other and to previous proposals, and what further challenges remain.  

This paper addresses these issues by analyzing and experimentally evaluating state-of-the-art PM range indexes built for real PM. 
We find that newer designs inherited past techniques with new improvements, but do not necessarily outperform pre-Optane era proposals. 
Moreover, PM indexes are often also very competitive or even outperform indexes tailored for DRAM, highlighting the potential of using a unified design for both PM and DRAM. 
Functionality-wise, these indexes still lack good support for variable-length keys and handling NUMA effect. 
Based on our findings, we distill new design principles and highlight future directions.  
\end{abstract}

\maketitle

\newcommand\vldbavailabilityurl{https://github.com/sfu-dis/pibench-ep2}

\keywords{\thepaperkeywords}

\section{Introduction}
\label{sec:intro}
The recently commercialized persistent memory (PM) devices, represented by Intel Optane Persistent Memory (Optane PMem)~\cite{DCPMM}, deliver persistence, high capacity, lower cost and fast speed on the memory bus. 
There have been many PM-inspired (re)designs in data-intensive systems~\cite{Viper,Spitfire,Zen,Hymem,SOFORT,Arulraj2015,FEDB}. 
In particular, much exciting progress has been made on devising single-level persistent OLTP indexes that directly operate and store data on PM without involving the storage stack, even before real PM devices became available~\cite{Hwang2018,Chen2015,HiKV,NV-Tree,WORT,Venkataraman2011,PMwCAS,DBPCM,FPTree,BzTree}. 
As we have shown in the past~\cite{PiBench}, although these pre-Optane proposals do not perform as expected (e.g., much slower) on real PMem devices due to inaccurate assumptions and emulation, several building blocks (unsorted nodes, fingerprinting, hybrid DRAM-PM structures) have proved useful for devising new indexes on PM.

The availability of real PM devices has further enabled a new breed of PM-based indexes~\cite{Dash,roart,utree,dptree,lbtree,pactree}, which are tailor-made for Intel Optane PMem. 
Although they appear/claim to perform better than pre-Optane proposals on real PM, it remains unclear (1) \textit{how they compare against each other}, as these new indexes appeared roughly concurrently and/or were proposed by different research communities (e.g., VLDB/SIGMOD vs. SOSP/OSDI), (2) \textit{how they are different from (or similar to) the pre-Optane proposals} (i.e., what ``legacy'' has pre-Optane proposals left for the new breed?), and (3) \textit{what further challenges and opportunities remain in this area}.
The goal of this paper is to answer these questions, which is the key to pushing actual adoption of these new indexes and PM-based systems in general. 
We do so by conducting a comprehensive evaluation of five state-of-the-art PM indexes, including \dptree~\cite{dptree}, \utree~\cite{utree}, \lbtree~\cite{lbtree}, \roart~\cite{roart} and \pactree~\cite{pactree}. 
We also include as a reference FPTree~\cite{FPTree}, a representative and arguably the best-performing design from the pre-Optane era~\cite{PiBench}.

This work can be seen as a sequel to the ``first episode'' of our past work which benchmarked and distilled the aforementioned useful building blocks from pre-Optane era PM indexes~\cite{PiBench}, but is not a mere repeat of what was done before with newer indexes.
On the one hand, we follow the similar methodology and focus on representative range indexes. 
On the other hand, in this paper we give a snapshot of the latest state of this area and highlight new findings, observations and perspectives that were often omitted in the past, especially variable-length key support, NUMA-awareness and new potential impact of existing PM indexes on future work. 
We summarize our findings below. 
\circled{1} The key optimization target remains to be reducing read and (more often) write operations on PM to preserve PM bandwidth (and thus achieving higher performance). 
\circled{2} As we predicted in past work, several pre-Optane techniques are effective on real PM hardware and widely adopted by new proposals, including fingerprinting~\cite{FPTree}, unsorted nodes and leveraging DRAM. 
\circled{3} Although modern PM-enabled servers are often multi-socket, most state-of-the-art PM indexes are still only optimized for single-socket machines and consciously avoided NUMA effect in their experimental evaluations; 
handling NUMA effect remains an unsolved problem. 
\circled{4} PM programming infrastructure (e.g., PM allocators and runtime) is still far from ideal in many cases and requires further improvements. 
\circled{5} Finally and perhaps most profoundly, for the first time, we observe that techniques employed by PM indexes can also be useful in devising high-performance volatile in-memory indexes. 
Surprisingly, when running on DRAM without the extra fences and cacheline flushes, under certain workloads a PM index can match or even outperform well-tuned indexes specifically designed for DRAM, such as HOT~\cite{HOT} and Masstree~\cite{Masstree}. 
This highlights the potential of unifying persistent and volatile indexing to simplify the design and implementation of future systems. 

The remaining sections expand on more details and insights. 
We give the necessary background on PM in Section~\ref{sec:bg}, followed by a review of pre-Optane PM range indexes in Section~\ref{sec:pre-optane}. 
Sections~\ref{sec:new}--\ref{sec:analyze-new} then survey and analyze state-of-the-art PM range indexes. 
Sections~\ref{sec:eval}--\ref{sec:observe} present our experimental results and observations. 
We present an outlook of future indexes in Section~\ref{sec:next}. 
Section~\ref{sec:related} covers related work and Section~\ref{sec:conclusion} concludes this paper. 
We have open-sourced our code and results at \url{\vldbavailabilityurl}.

\section{Persistent Memory}
\label{sec:bg}
We briefly review the properties of PM (in particular Intel Optane PMem\footnote{Also known as Optane DC Persistent Memory Module (DCPMM)~\cite{DCPMMBrief}.}) and its implications on software; readers already familiar with the literature may skim and fast forward to Section~\ref{sec:pre-optane}. 

Collectively, ``PM'' refers to a class of devices that offer byte-addressability (like DRAM), high capacity and persistence (like SSDs) on the memory bus. 
They can be built using various materials and techniques, such as PCM~\cite{PCM}, STT-RAM~\cite{STT-RAM} and memristor~\cite{Memristor}. 
However, only the 3D XPoint~\cite{Intel3DXP} based Optane PMem so far has delivered high capacity promised by the PM vision.
Thus, we target Optane PMem in this paper. 
Optane PMem's performance generally fall between DRAM and SSDs.\footnote{Exceptions apply under certain workloads and hardware configurations~\cite{SSDStrikingBack}.} 
The first generation (100 series) exhibits $\sim$300ns latency and $\sim$5--40GB/s bandwidth depending on the access pattern, and with sequential/read accesses being faster than random/write accesses. 
The latest 200 series further gives $\sim$30\% higher bandwidth~\cite{Optane200}.

Although PM offers persistence, 
it is still behind multiple levels of CPU caches. 
Software normally goes through CPU caches to access data stored on PM using \load and \store instructions, and explicitly issues cacheline flush instructions (e.g., \CLWB and \CLFLUSHOPT) and fences to ensure data is correctly persisted~\cite{IntelManual}. 
However, PM-capable platforms use asynchronous DRAM refresh (ADR) to guarantee data flushed from the CPU caches will first land on the CPU's write buffer, which is power failure protected. 
As a result, a PM write is considered complete once the data is forced to the ADR domain, without necessarily having arrived at physical PM media. 
More recent platforms further feature enhanced ADR (eADR) which also protects the CPU cache, effectively providing durable CPU caches and potentially sparing the need of cacheline flush instructions (although fences are still needed for correct ordering). 
eADR is only available for the more recent 200-series Optane PMem and Skylake platforms~\cite{IntelManual}. 
Our experiments are based on the 100-series PMem and Cascade Lake platform without eADR (detailed setups in Section~\ref{sec:eval}). 
However, we do not expect our conclusions to change based on recent results obtained from evaluating the impact of eADR~\cite{PMIdio}. 
We leave it as future work to document eADR's detailed behavior and impact on PM range indexes.

Optane PMem can operate under the Memory, App Direct, or Dual modes~\cite{IntelManual}. 
The Memory mode uses the system's DRAM as a hardware-controlled cache and presents bigger but volatile memory. 
The App Direct mode enables persistence, thus allowing building persistent indexes. 
The Dual mode combines both by allowing part of PM to be configured for the Memory or App Direct mode. 
Same as other work, we focus on the App Direct mode as it gives software the flexibility to use DRAM and PM as needed.

\begin{table*}[t]
\caption{Design choices of pre-Optane PM index proposals (\wbtree, \bztree, \fptree and \nvtree)~\cite{PiBench}. 
Common building blocks such as using DRAM to store reconstructible data (e.g., inner nodes), using unsorted nodes and fingerprinting have proved to be useful on real Optane PMem. 
The impact of variable-length keys, PM allocator and NUMA effect were largely unconsidered. 
}
\label{tbl:indexes}
\begin{tabular}{@{}m{1.9cm}m{3.2cm}m{3.4cm}m{2.8cm}m{1cm}m{2.3cm}m{1cm}@{}}
\toprule
\textbf{}        & \textbf{Architecture} & \textbf{Node Structure} & \textbf{Concurrency} & \textbf{Var. Keys} & \textbf{PM Allocator} & \textbf{NUMA-Aware} \\ \midrule
\textbf{\wbtree}~\cite{Chen2015}  & B+-tree; PM-only &  Unsorted; indirection  & Single-threaded & Pointer                              & Emulation/PMDK & No \\
\rowcolor[HTML]{EFEFEF} 
\textbf{\bztree}~\cite{BzTree}  & B+-tree; PM-only &  Partially unsorted leaf & Lock-free with Persistent MwCAS~\cite{PMwCAS}  & Inlined & Emulation/PMDK & No \\
\textbf{\fptree}~\cite{FPTree}  & B+-tree; DRAM (inner) + PM (leaf) &  Unsorted leaf; fingerprints & HTM (inner) + locking (leaf) & Pointer & Customized/PMDK & No \\ 
\rowcolor[HTML]{EFEFEF} 
\textbf{\nvtree}~\cite{NV-Tree} & B+-tree; PM-only or optionally DRAM+PM & Unsorted leaf; inconsistent inner nodes & Locking & Pointer & Emulation/PMDK & No \\\bottomrule
\end{tabular}
\end{table*}

\section{Previously on PM Range Indexes}
\label{sec:pre-optane}
Indexing has received much attention even before real PM was available~\cite{FPTree,BzTree,NV-Tree,Chen2015,Venkataraman2011,DBPCM,WORT,PMwCAS,Hwang2018}. 
This section reviews pre-Optane designs to set the stage for our new evaluations. 

\subsection{Early Assumptions in the Pre-Optane Era}
Due to the lack of real devices, early proposals had to ``guesstimate'' the properties of PM based on prototypes and simulations~\cite{PCM,ZhouPCM,LeePCM}.\footnote{Various materials can be used to build PM, yet they may perform differently. This partially forced past work to make general assumptions for potentially wider applicability.} 
Common assumptions included (1) limited write endurance, (2) 3--5$\times$ higher latency but similar bandwidth to DRAM's~\cite{NVM-DLog,Pelley2014}, (3) writes slower than reads, (4) 8-byte atomic PM writes~\cite{BPFS}, and (5) volatile CPU caches and reorderings by the CPU. 
Out of these, the assumption on bandwidth turned out to be inaccurate: Optane PMem's lower-than-DRAM bandwidth is a major factor that limits performance~\cite{PiBench}. 
The speed gap between sequential and random accesses was also largely left out. 
Endurance so far has not been a major issue for Optane PMem which warrants virtually unlimited endurance during the usual replacement cycle of 3--5 years~\cite{DCPMMEnduranceReport,DCPMMBrief}. 

Nevertheless, these assumptions rightfully suggested that classic in-memory indexes would not guarantee correctness (customized recovery protocols are necessary), nor perform well on PM. 
Concurrency control must be carefully considered given PM's higher latency. 
These led to the development of numerous PM indexes even before actual PM hardware was available. 
The key is reducing PM read/write operations and avoiding unnecessary cacheline flushes and fences to both improve performance and reduce wear. 

\subsection{Presumed Designs and Building Blocks}
\label{subsec:pre-optane-designs}
To reduce PM accesses, several building blocks have been proposed around re-designing the tree architecture, node structure and concurrency control. 
Now we discuss them in the context of previously evaluated PM indexes (\wbtree~\cite{Chen2015}, \bztree~\cite{BzTree}, \fptree~\cite{FPTree} and \nvtree~\cite{NV-Tree}). 
Table~\ref{tbl:indexes} lists their main design choices. 
The first three dimensions (architecture, node structure and concurrency) received the most attention in the past; 
in this work, we also consider variable-length keys, PM management and NUMA-awareness. 

\textbf{Tree Architecture: PM-Only $\rightarrow$ DRAM-PM Hybrid.}
Most pre-Optane proposals adapt in-memory B+-trees. 
Some (e.g., \wbtree and \bztree) place the entire tree on PM to allow instant recovery. 
But doing so leads to much slower traversal speed compared to DRAM indexes. 
A key contribution by \fptree and \nvtree was to leverage the fact that B+-tree's inner nodes only guide search traffic and are reconstructible using leaf nodes. 
This allows improved traversal speed by loosening the consistency requirements for inner nodes, by omitting flushes/fences~\cite{NV-Tree} and/or placing all the inner nodes in DRAM~\cite{FPTree}. 
Upon restart, the inner nodes can be rebuilt using B+-tree bulk loading algorithms. 
The downside is recovery time can scale with data size, sacrificing instant recovery. 

\textbf{Node Structure: Sorted $\rightarrow$ Unsorted.}
Traditional B+-trees keep keys sorted for fast binary search. 
The drawback of inheriting this design on PM is insertions may shift keys, causing excessive PM reads and writes, while having the risk of incomplete updates upon hardware failure. 
Moreover, binary search becomes less (or not at all) beneficial with small nodes commonly used by in-memory B+-trees.
A popular solution is to keep nodes unsorted and use a linear scan to retrieve the target key. 
Most of the surveyed pre-Optane indexes adopted this technique. 
To mitigate the impact of linear search, BzTree periodically consolidates nodes to become sorted. 
FPTree proposed fingerprinting which maintains one-byte hashes of the keys in the node and a lookup starts by checking the fingerprints. 
Only the keys with matched fingerprints will be further examined. 
This greatly reduces PM accesses, especially for negative search where the key does not exist. 
Another approach is to keep an indirection array~\cite{Chen2015} that stores sorted index positions of keys, allowing binary search using the indirection array. 

\textbf{Concurrency Control: Pessimistic $\rightarrow$ Optimistic.}
PM indexes often prefer lightweight concurrency control over pessimistic lock coupling~\cite{CowBook}. 
\fptree uses separate strategies for inner and leaf nodes: For the former it leverages hardware transactional memory (HTM) to reduce traversal costs, and for the latter it uses traditional locking because inserting into PM-resident leaf nodes may involve cacheline flushes, which in turn will abort HTM transactions. 
\bztree uses lock-free multi-word compare-and-swap (PMwCAS)~\cite{PMwCAS} that can atomically modify multiple 8-byte words. 
Without pessimistic locking, \fptree scales to high core count but exhibits high tail latency and low throughput under contention, due to HTM's inherent limitations~\cite{PiBench}. 
\fptree delegates the detailed HTM-locking interactions 
to Intel TBB which uses a fast-path that uses HTM and a slow-path that serves as a fallback when HTM abort rate becomes higher than a threshold (10 by default).\footnote{Details at \url{https://github.com/oneapi-src/oneTBB/blob/v2021.5.0/src/tbb/rtm\_mutex.cpp\#L33}. Our evaluation (Section~\ref{sec:eval}) sets the threshold to 256 for better performance.} 

\subsection{Functionality and PM Management}
\label{subsec:pre-optane-func}
Now we turn to the remaining three dimensions under consideration in Table~\ref{tbl:indexes}: support for variable-length keys, NUMA awareness and PM programming infrastructure support.

Most pre-Optane proposals~\cite{FPTree,Chen2015,NV-Tree} focused on handling fixed-length, 8-byte keys, so were past evaluation efforts. 
Variable-length keys are usually supported using pointers to keys stored in the (persistent) heap. 
Some proposals differentiate pointers from inlined values by designating a special ``type'' bit in the 8-byte key area, effectively limiting the maximum length of keys to be 63 bits. 
Others, e.g., \fptree, require compile-time customization by specifying whether keys are 8-byte integers or pointers to support full 64-bit keys. 
Pointers are used in case both types are required. 
As a result, accessing shorter keys $\le$ 8 bytes is also prone to cache misses caused by pointer chasing. 
Out of the four indexes, only \bztree provides inlined support for variable-length keys with slotted pages. 

None of the surveyed pre-Optane indexes handle NUMA effect. 
Most of them also do not have a well thought-out design for managing PM. 
\fptree uses PMDK~\cite{PMDK}, the current de facto standard PM library, to avoid issues such as PM leaks. 
For better performance, \fptree has to use customized slabs (large chunks allocated from PMDK allocator) to amortize PM allocation cost. 
Still, our previous work~\cite{PiBench} has identified PM allocation in all these indexes as a main bottleneck that should be removed by future designs.

\section{State-of-the-Art PM Range Indexes}
\label{sec:new}
We survey five representatives of recent PM indexes optimized for Optane PMem: \lbtree~\cite{lbtree}, \dptree~\cite{dptree}, \utree~\cite{utree}, \roart~\cite{roart} and \pactree~\cite{pactree}.
They can be categorized as B+-tree based, trie based and hybrid which makes use of both B+-trees and tries.\footnote{In addition to traditional indexes, learned indexes~\cite{LearnedIndex} are also being adapted for PM~\cite{APEX}. However, they are still in very early stage. We thus leave it as future work to evaluate them to avoid pre-mature conclusions.} 

\subsection{B+-Tree based: \lbtree and \utree}
\label{subsec:btree}
We survey two representative B+-tree based PM range indexes, \lbtree~\cite{lbtree} and \utree~\cite{utree}, which represent designs that mainly optimize for high throughput and low tail latency, respectively.

\textbf{\lbtree.} As evaluated by previous work~\cite{PiBench,UCSDMeasurement,UCSDGuide}, Optane PMem uses 256-byte granularity internally: accesses smaller than 256 bytes will still incur 256-byte of traffic to/from the physical media.
It then becomes important to (1) coordinate PM accesses in 256-byte granularity and (2) reduce unnecessary PM accesses (to save PM bandwidth)~\cite{UCSDGuide,UCSDMeasurement}.
To reach these goals, as shown in Tables~\ref{tbl:indexes} and \ref{tbl:new-indexes}, \lbtree starts with several useful techniques proposed by \fptree: fingerprinting, DRAM+PM architecture and optimistic concurrency with HTM and locking.
Compared to \fptree, it uses hand-rolled RTM transactions instead of TBB. 
We highlight the impact of this design decision later in Section~\ref{sec:eval}.

\begin{figure}[t]
	\centering
	\includegraphics[width=\columnwidth]{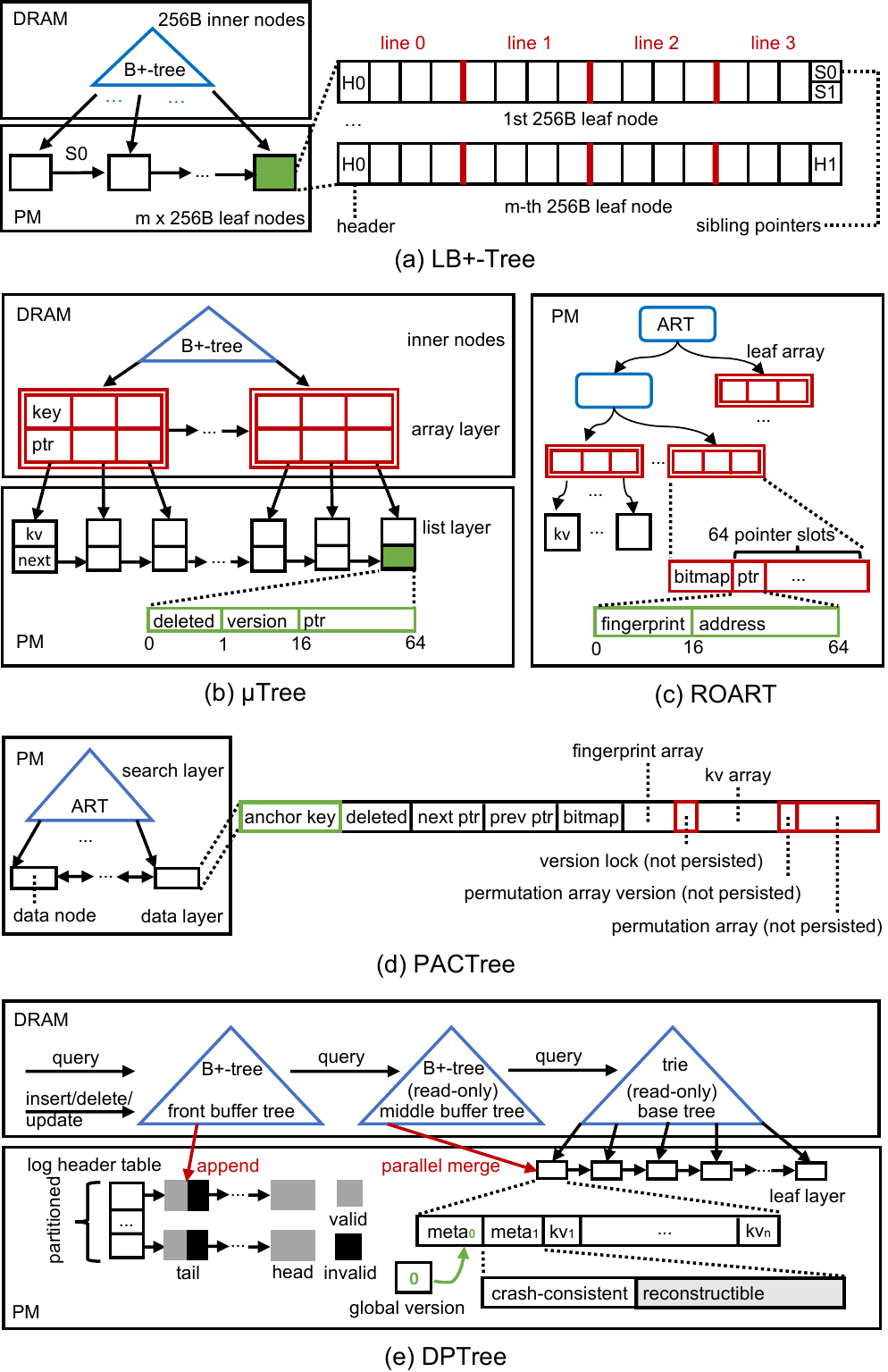}
	\caption{Architecture of five state-of-the-art PM indexes.}
	\label{fig:trees}
\end{figure}

On top of these existing techniques, \lbtree proposes new ones to further optimize PM accesses.
First, node sizes are set to align with and be multiples of 256 bytes for better CPU cache and PM utilization.
Second, to minimize PM writes during inserts, \lbtree proposes entry moving to bound the number of cacheline writes per insert.
As Figure~\ref{fig:trees}(a) shows, a 256-byte leaf node is divided into four 64-byte cachelines.
Upon insert, \lbtree first attempts to insert the record into \texttt{Line 0} if an empty slot is available in it.
Since the header is also in \texttt{Line 0}, only one cacheline write to PM is needed to persist both the record and header. 
Otherwise, \lbtree moves data from \texttt{Line 0} to another line where the new entry is inserted.
This proactively spares empty slots in the first cacheline, which will reduce flush operations incurred by future inserts.

Write-ahead logging (WAL) is widely used for crash recovery ~\cite{PMDK}, but incurs additional PM writes.
\lbtree disposes of WAL with logless node splits by storing two \texttt{sibling} pointers in each leaf node and uses an alt bit in node header to indicate the valid pointer.
Upon split, a new leaf node is first allocated and tracked by the unused pointer, followed by a redistribution of entries to the last two cachelines of the new node.
Then the alt bit and bitmap in the original node are updated using an 8-byte atomic PM write
to ensure the tree is always in a consistent state even across failures.

\textbf{\utree.}
Different from most PM indexes, \utree mainly optimizes for tail latency caused by PM's high latency.
It places B+-tree inner nodes in DRAM, but redesigns leaf nodes to use both DRAM and PM.
As shown in Figure~\ref{fig:trees}(b), leaf nodes in \utree consist of two layers: an array layer and a list layer.
The former consists of traditional B+-tree nodes which store pointers to list layer nodes; the latter is a PM-resident singly-linked list where each node stores a key-value pair.
\utree uses optimistic locking~\cite{ARTOLC} for the DRAM-resident B+-tree and manages the PM-resident linked list in a lock-free manner.
To insert a record, the thread first traverses the in-DRAM B+-tree optimistically without holding any locks, and then inserts a node representing the key-value pair in the linked list using the compare-and-swap (\CAS) instruction~\cite{PracticalLockFree,HarrisLockFreeLinkedList}.
It then acquires the lock in the corresponding leaf node in the B+-tree to insert the key, which may trigger splits that will in turn acquire locks from the bottom up in DRAM.
This way, B+-tree structural modification operations (SMOs) and actual key-value inserts/removals (which incur PM flushes in the list layer) are decoupled, and multiple threads inserting into the same leaf node could proceed in parallel in the list layer. 
This could allow more concurrency and removes flushes from the critical path, thus reducing tail latency.

\subsection{Trie based: \roart and \pactree}
\label{subsec:trie}
Newer designs have also explored tries on PM.
They are mostly PM-only (instead of DRAM-PM hybrids) and based on ART~\cite{ART}. 
We survey two representative proposals, \roart~\cite{roart} and \pactree~\cite{pactree}.

\textbf{\roart.}
Based on ART, \roart optimizes for range scans, which are known to be a weak point of tries~\cite{roart}. 
Figure~\ref{fig:trees}(c) shows the overall architecture of \roart. 
It is purely on PM but selectively persists metadata entries and reconstructs the inconsistent ones upon recovery (similar to \nvtree's selective consistency~\cite{NV-Tree}). 
To optimize scan, \roart compacts small sub-trees ($<$ 64 entries) into leaf arrays that store pointers to records. 
This reduces pointer chasing overhead during scan and makes the index shallower.
The downside is splits become expensive, as all the keys in a leaf array are divided into subsets based on the first differentiating byte, followed by a node allocation for each subset.
This requires more PM writes and fences, and puts higher pressure on the PM allocator.
To this end, \roart reduces the number of fences by relaxing the order of split steps and using a depth field to detect and resolve inconsistency.
It also proposes delayed check memory management (DCMM) to cope with the high PM allocation demand. 
DCMM performs fast allocations using thread-local pools but delays garbage collection by traversing the entire index at a later time, potentially increasing PM usage. 
Since \roart is fully in PM, it supports true instant recovery, and natively supports variable-length keys without extra pointer chasing using its trie-based design.

\begin{table*}[t]
\caption{Main design choices of state-of-the-art PM range indexes (\lbtree, \dptree, \utree, \roart and \pactree).
They base on Intel Optane PMem and inherit designs from pre-Optane proposals, by using DRAM (sometimes more aggressively), lightweight concurrency control and unsorted nodes.
Some also advocate customized PM allocators to reduce PM accesses.
Variable-length keys and NUMA-awareness are still less considered.
}
\label{tbl:new-indexes}
\begin{tabular}{@{}m{1.9cm}m{3.2cm}m{3.4cm}m{2.8cm}m{1cm}m{2.3cm}m{1cm}@{}}
\toprule
\textbf{}        & \textbf{Architecture} & \textbf{Node Structure} & \textbf{Concurrency} & \textbf{Var. Keys} & \textbf{PM Allocator} & \textbf{NUMA-Aware} \\ \midrule
\textbf{\lbtree}~\cite{lbtree}  &  B+-tree; DRAM (inner) + PM (leaf) &  Unsorted leaf; fingerprints; extra metadata& HTM (inner) + locking (leaf) &  Pointer  & Customized/PMDK & No \\
\rowcolor[HTML]{EFEFEF}
\textbf{\utree}~\cite{utree}   & B+-tree; DRAM (B+-tree) + PM (linked list) &  Sorted & Locking (array layer) + lock-free (list layer) & Pointer & PMDK & No \\
\textbf{\dptree}~\cite{dptree}  & Hybrid; DRAM (B+-tree, trie inner) + PM (trie leaf) &  Unsorted leaf; fingerprints; indirection; extra metadata & Optimistic lock~\cite{ARTOLC} + async. updates &  Pointer & PMDK & No\\
\rowcolor[HTML]{EFEFEF}
\textbf{\roart}~\cite{roart}   & Trie; PM-only & B+-tree like unsorted leaf; fingerprints & ROWEX~\cite{ARTOLC} & Inlined & Customized/PMDK & No\\
\textbf{\pactree}~\cite{pactree} & Trie; PM-only or optionally DRAM+PM & Unsorted leaf; fingerprints; indirection & Optimistic lock~\cite{ARTOLC} + async. update & Inlined & Customized/PMDK & Yes \\ \bottomrule
\end{tabular}
\end{table*}

\textbf{\pactree.}
\pactree is a PM-only two-layer persistent trie with B+-tree styled leaf nodes.
As shown in Figure~\ref{fig:trees}(d), \pactree consists of a search layer and a data layer.
The search layer is a durable trie based on concurrent ART that uses read-optimized write exclusion (ROWEX)~\cite{ARTOLC}.
The data layer is a doubly-linked list of B+-tree leaf nodes, each of which contains 64 key-value pairs and an anchor key to indicate the smallest key in the node.
\pactree stores fingerprints and indirection arrays in leaf nodes to facilitate search and scan, but they are not persisted to reduce PM writes. 
Upon split, the target leaf node in the data layer is first locked, and then a log entry is written to a per-thread SMO log in PM.
The thread then splits the leaf node and commits without modifying inner nodes in the search layer.
A background thread will then finish the remaining SMO in the search layer.
This allows worker threads to commit early right after modifying leaf nodes.
However, it also creates inconsistencies between the search and data layers.
Thus, query threads may need to perform a ``last-mile'' search after arriving at the data layer to find the correct leaf node using anchor keys.
\pactree is the only index that mitigates NUMA effect.
It uses separate pools for the search layer, data layer and logs in each NUMA node.
It also advocates the use of snooping-based CPU coherence protocols (instead of the default directory-based protocol on most platforms) to avoid poor performance when PM accesses cross NUMA boundaries.

\subsection{Hybrid: \dptree}
\label{subsec:hybrid}
As shown in Figure~\ref{fig:trees}(e), \dptree is a PM-DRAM hybrid index that combines up to two B+-trees (a \textit{front} and a \textit{middle} buffer tree) in DRAM, a trie (\textit{base} tree) that places inner nodes in DRAM and leaf nodes in PM.
To search for a key, \dptree first visits the front buffer tree.
If the target key is not found, the middle buffer tree (if exists) will be further searched.
If the key does not exist in the buffer trees, the base tree will have to be searched.
To reduce unnecessary traversals, \dptree maintains a bloom filter per buffer tree.
For range queries, \dptree has to search and merge results from all the trees.
When the size ratio between the front buffer tree and base tree reaches a pre-defined threshold, \dptree creates a new front buffer tree and turns the previous front buffer tree into a middle buffer tree.
Tree merge operations are triggered when the size ratio between the front buffer tree and the base tree reaches a pre-defined threshold, and are performed by background threads.
Using the version number and extra set of metadata, \dptree ensures changes are invisible to concurrent queries when a merge is in progress.
After merging, the middle buffer tree is destroyed and the inner nodes of the base tree (ART) are rebuilt.
Then the global version bit is flipped to expose changes to incoming requests. 
\dptree also uses selective metadata persistence with reconstructible metadata (e.g., record count and fingerprints) in DRAM.

\section{Analyzing the State-of-the-Art}
\label{sec:analyze-new}
With the high-level designs laid out, now we analyze the new PM indexes in detail and distill common building blocks which can be useful for future PM indexes.
We analyze them from the six dimensions in Table~\ref{tbl:new-indexes}, followed by empirical evaluation in Section~\ref{sec:eval}.

\subsection{Index Architecture}
New PM indexes often inherit the PM+DRAM architecture with new optimizations, and consider tries and hybrid structures.

\textbf{(More Extensive) Use of DRAM.}
New PM indexes based on B+-tree and hybrid structures (\lbtree, \dptree and \utree) continue to use DRAM to store part of the index.
We also observe more aggressive use of DRAM, e.g., \dptree and \utree place entire tree structures in DRAM to get more performance gains.
The tradeoffs are (1) longer recovery time, (2) more complex programming and (3) higher DRAM consumption which we quantify in Section~\ref{sec:eval}.

\textbf{Beyond B+-Trees and Monolithic Indexes.}
New PM indexes also adapt tries (\pactree and \roart), but they default to pure PM designs, potentially leading to suboptimal performance but preserving instant recovery.
Notably, \pactree stores keys in both inner and leaf nodes, but it is possible to place inner nodes in DRAM to improve performance.
We expect to see more B+-tree based PM indexes utilize DRAM for faster access/write speed, as PM servers will still feature DRAM in the foreseeable future.\footnote{DRAM must be present for PMem to work, even if the software does not need it~\cite{SSDStrikingBack}.}
As a major departure from just adapting one type of data structure, \dptree combines B+-trees and tries. 
When it comes to node structure, new indexes are also introducing new designs that no longer use pure trie or B+-tree nodes, which we highlight next.

\subsection{Node Structure}
New PM indexes base their designs on Optane PMem with node alignment of 256 bytes to reduce unnecessary PM accesses.
Several pre-Optane designs---fingerprinting, unsorted (leaf) nodes and selective consistency for metadata---continue to be used by new PM indexes.
But they are further optimized with new techniques. 

\textbf{Fingerprinting on Steroids.}
As PM accesses are slow, fingerprinting becomes the most popular approach used by four out of the five new PM indexes.
Meanwhile, new techniques are introduced to better store and use fingerprints.
\lbtree uses SIMD instructions to compare up to 64 one-byte fingerprints in one instruction.
\roart embeds a two-byte fingerprint inside pointers to key-value pairs, minimizing pointer chasing at the leaf level.

\textbf{Extra and Selectively Persisted Metadata.}
\lbtree and \dptree both use an extra set of metadata per leaf node to avoid logging (thus reducing PM writes). 
For \lbtree, this allows it to achieve logless split.
\dptree uses the extra metadata to track PM allocations and hide incomplete changes during tree merge operations.

Not all metadata entries have to be persisted when modified.
For instance, version locks in \pactree leaf nodes are only meaningful at runtime;
fingerprints and indirection arrays can be rebuilt during recovery (\dptree and \pactree) or on demand at runtime by query threads (\roart).
Keeping them volatile can significantly reduce PM writes and improve performance, at the cost of slower recovery.

\textbf{Hybrid Leaf Nodes.}
All the surveyed PM indexes that adopt trie (\dptree, \roart and \pactree) use B+-tree like leaf nodes where a node stores multiple records to reduce insert overhead, as the leaf node is only split when it is full. 
This design also reduces pressure on the PM allocator, as trie-based indexes typically incur more frequent allocations with varying node sizes compared to B+-trees.
Scan performance is also improved with less pointer chasing.
Different from other proposals, \utree introduces a linked list layer in PM, making its ``leaf node'' logical.
Although this design decouples SMOs and data movement to potentially enable more parallelism, it adds more overhead to scans due to more pointer chasing.

\subsection{Concurrency Control}
All the surveyed new PM indexes use optimistic concurrency control.
They optimize traversals using lock-free read or HTM for inner nodes;
locks are only acquired at the leaf level and/or as needed in inner nodes to reduce PM writes.
Further, they tend to use background threads for SMOs (e.g., \pactree and \dptree). 
The benefit of offloading SMOs to the background is potentially lower latency for index operations.
However, it can be tricky to determine the appropriate number of background threads.
Also, with a given CPU budget (e.g., in the cloud), the machine may not have enough resources to spare for the background threads, which may then fall behind the foreground threads and affect the overall progress.

\subsection{Functionality and PM Management}
As Table~\ref{tbl:new-indexes} lists, among the new PM indexes, only trie-based \roart and \pactree natively support variable-length keys; the others follow pre-Optane proposals to use pointers to keys. 
With real hardware and libraries like PMDK, all the indexes have taken into account PM management issues (e.g., avoiding persistent leaks and optimizing allocation performance).
However, many need a customized allocator for performance reasons. 
Finally, only \pactree is designed to mitigate NUMA effect.
Other indexes would have to use general-purpose approaches~\cite{NAP} that can be applied on any PM index, but they come with limitations (e.g., focus on certain workloads).
Such facts indicate that in terms of functionality, new PM indexes have been mainly sticking with the status quo.

\section{Evaluating the State-of-the-Art} 
\label{sec:eval}
Now we empirically evaluate new PM indexes and compare them with \fptree, the best-performing pre-Optane PM range index. 

\subsection{Experimental Setup}
We run experiments on a 40-core (80-hyperthread) server equipped with two Intel Xeon Gold 6242R CPUs clocked at 3.10 GHz with 36MB of cache. 
The server is fully populated with 12$\times$32GB DRAM DIMMs (384GB in total) and 12$\times$128GB Optane PMem DIMMs (1.5TB in total) 
for maximum bandwidth. 
Both DRAM and PMem run at 2666MT/s.\footnote{DRAM has to be clocked down from 3200MT/s to 2666MT/s for PMem to work~\cite{DCPMMBrief}.}
The server runs Arch Linux with kernel 5.14.9. 
All the code is compiled using GCC 11.1 with all the optimizations. 
Unless otherwise specified, we use PMDK~\cite{PMDK}/\jemalloc~\cite{jemalloc} for PM/DRAM allocations. 

\textbf{Benchmarking Framework.}
We use \pibench~\cite{PiBench}, a unified framework for benchmarking PM indexes, to stress test the indexes. 
\pibench generates and issues synthetic workloads of given distributions that consist of user-specified operations (lookup/insert/update/delete/scan). 
It requires each index implement a set of common interfaces, by extending an abstract C++ class. 
We create a wrapper for each index that uses \pibench's interfaces to invoke the index's internal operations. 
The wrapper is compiled as a shared library and loaded into \pibench's address space at runtime. 

\textbf{Metrics and Workloads.}
We measure throughput (operations per second) and latency at various thread counts. 
Using \pibench, we collect statistics such as cache misses and bandwidth to aid analysis.
We test both fixed-length (8-byte) integer and variable-length keys, following the setups used by previous work~\cite{PiBench}. 
For point queries we use keys chosen randomly under uniform or skewed distributions; 
for range scans, we uniform randomly choose a start key $K$ and scan 100 records following $K$. 
We prefill each index with 100 million key-value pairs, after which we start to run individual and mixes of index operations. 
The results across runs vary little and so we report the average of three 10-second runs.

\subsection{Index Implementations and Parameters}
For all the indexes (except \fptree which we had to implement), we use the original authors' code obtained from their public repositories. 
We use parameters that lead to the best performance (described below) and make necessary changes to each index for correctness, functionality and fairness. 

\textbf{\lbtree}. 
(1) The original insert and delete functions do not guarantee persistence; 
we followed the authors' suggestions to fix them.\footnote{Details at \url{https://github.com/schencoding/lbtree/issues/2}.}
(2) We implemented range scan (similar to \fptree) and update with locking in leaf nodes.
(3) We found the PMDK allocator can provide sufficient performance after tuning PMDK parameters and allocating 256-byte aligned leaf nodes, so we also use PMDK for \lbtree.
The inner/leaf nodes contain 15/14 entries (256-byte). 

\textbf{\utree}. 
Instead of using chunk-based allocation, the original code in fact uses PMDK's \texttt{POBJ\_ZALLOC}. 
We changed it to use \texttt{pmemobj\_alloc} for much better performance. 
We were not able to verify the correctness of multi-threaded inserts (with keys missing after successful inserts), 
so we only include \utree in single-threaded experiments. 
Inner/leaf node sizes are both set to 29 entries. 

\textbf{\roart}. 
The open-source code of \roart supports both PMDK and its customized DCMM allocator~\cite{roart}; we present numbers under both allocators. 
Leaf node size is set to 64 entries.

\textbf{\pactree}. 
We use \pactree's own NUMA-aware PM allocation. 
For fair comparison, we pin the background and worker threads to the same CPU cores so that all the indexes use the same amount of CPU cores. 
Node size is set to 64 entries. 
 
\textbf{\dptree}. 
The original code misses PM allocator support, so we ported it to use PMDK. 
We follow the original paper to use an equal number of worker and merge threads. 
For fair comparison, we also pin the merge and worker threads to the same cores (similar to \pactree's setup). 
Inner and leaf nodes contain 31 entries for buffer trees; 
the base tree uses 256-entry leaf nodes. 

\textbf{\fptree}. 
Since the original implementation is proprietary, we implemented \fptree by strictly following the paper.\footnote{Code available at \url{https://github.com/sfu-dis/fptree}.} 
We have verified that our implementation performs similarly to the original author's binary does. 
We follow past work's recommendations to set inner/leaf node sizes to 128/64 entries~\cite{PiBench,FPTree}. 

\subsection{Single-threaded Performance}
\label{subsec:single-thread-exp}
We begin with single-threaded experiments to avoid concurrency complicating our analysis. 
We run lookup/insert/update/scan operations under the uniform random distribution and report throughput. 

\begin{figure}[t]
\centering
\includegraphics[width=\columnwidth]{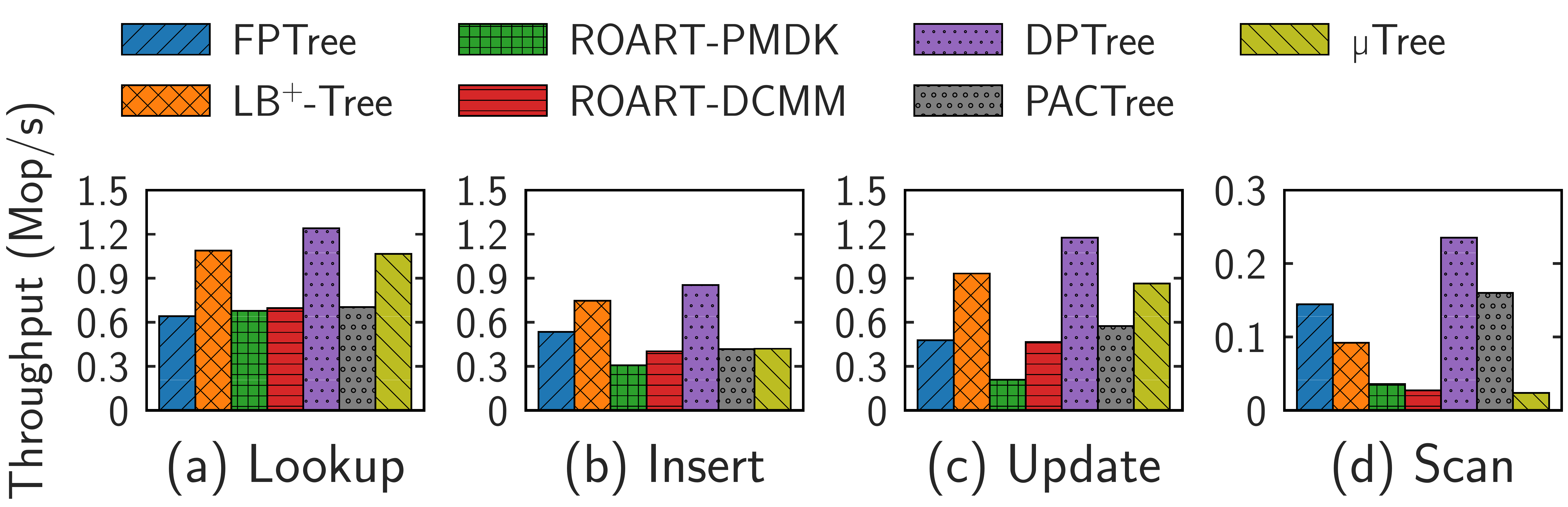}
\caption{Single-threaded throughput (uniform distribution).
Overall, \dptree and \lbtree perform the best. 
\fptree can be very competitive to (or even better than) newer indexes. }
\label{fig:pm-uniform-1t}
\end{figure}

\begin{figure*}[t]
	\centering
	\includegraphics[width=\linewidth]{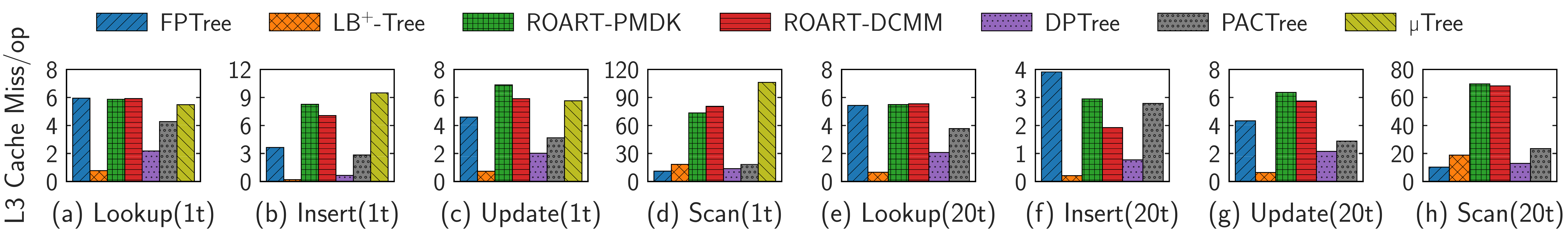}
	\caption{Last-level cache misses per operation under a single (a--d) and 20 threads (e--h).}
	\label{fig:cache_misses_pmem}
\end{figure*}

\textbf{Point Queries. } 
\lbtree, \dptree and \utree perform similarly for lookups in Figure~\ref{fig:pm-uniform-1t}(a). 
They are up to $\sim$2$\times$ faster than \fptree, \roart and \pactree, which also perform similarly. 
Overall, \dptree performs the best under single-threaded lookups, largely because of its extensive use of DRAM: 
if the search key is in the DRAM-resident buffer tree, the entire query can finish without ever accessing PM. 
If the base tree needs to be visited, only the search in leaf node will incur PM access, which is mitigated by binary search. 
The other two faster indexes (\lbtree and \utree) also benefit from placing inner nodes and the array leaf layer in DRAM. 
\lbtree also extensively uses SIMD instructions and prefetching, which as shown in Figure~\ref{fig:cache_misses_pmem}(a) drastically reduces cache misses and our factor analysis showed that prefetching alone improves performance by $\sim$10\%. 
For inserts, B+-tree based \fptree/\lbtree and hybrid \dptree are up to $\sim$2$\times$ faster than trie-based \roart and \pactree in Figure~\ref{fig:pm-uniform-1t}(b). 
\utree's use of linked lists in the leaf level cancels out some of B+-tree's advantages due to high cache miss rates in Figures~\ref{fig:cache_misses_pmem}(b--c). 
The performance of PM allocators is critical for \roart as for each update it needs to allocate a new PM block, and using its own DCMM can double the throughput for updates in Figure~\ref{fig:pm-uniform-1t}(c).
Compared to lookups, updates incur additional PM writes to update the payload, but will not trigger SMOs compared to inserts. 
As expected, the update performance of all indexes falls between their lookup and insert performance, but follows the trend of insert performance more closely because 
(1) PM exhibits higher write latency, and 
(2) similar to inserts, leaf-level locks can only be released after the new value is flushed, adding delays (although being a constant amount of overhead under a single thread). 

\textbf{Range Scans.}
As Figure~\ref{fig:pm-uniform-1t}(d) shows, range scan performance depends largely on the cost of scanning within and across leaf nodes, i.e., whether the nodes are sorted and they are big or small. 
Although \dptree needs to search multiple trees and combine results, it is still the fastest. 
\dptree's leaf nodes use indirection arrays, so the result can be returned directly without sorting, as oppose to trees that use unsorted nodes, e.g., \fptree.
\lbtree inherited a lot from \fptree, but is $\sim$30\% slower than \fptree for scans, because its leaf nodes are smaller (14 entries). 
To scan for the same number of records, compared to \fptree which uses 64-entry nodes, more leaf nodes have to be visited by \lbtree, causing more cache misses in Figure~\ref{fig:cache_misses_pmem}(d). 
This result highlights the tradeoff between hardware consciousness and optimization goals: 
using small nodes allows \lbtree to perform well in point queries, but can penalize scans.

\begin{figure*}[t]
	\centering
	\includegraphics[width=\linewidth]{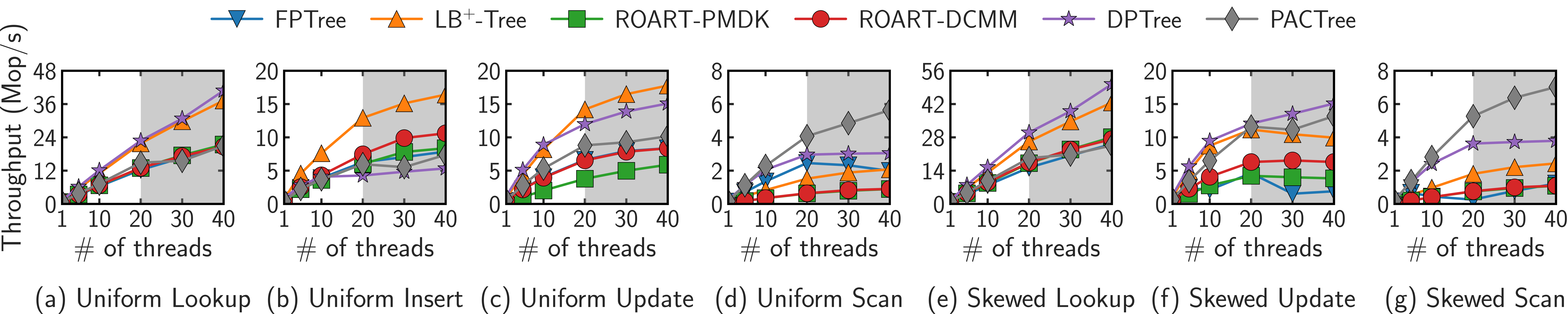}
	\caption{Throughput under uniform (a--d) and skewed (e--g, self-similar with 80\% accesses on 20\% of keys) distributions.} 
	\label{fig:pm-mt}
\end{figure*}

All the tested indexes except \roart first copy the scan results to an array and then optionally sort them before they are returned to the user. 
\roart, however, first returns an array of leaf pointers without copying. 
Moreover, to support variable-length keys, it stores in leaf nodes pointers to keys. 
This mandates the sorting pass to dereference pointers to keys, causing extra cache misses. 
We note that as an optimization, if key size is known, one may change \roart to copy keys first, which in our tests can double the performance at the cost of supporting variable-length keys. 
Finally, \utree exhibits low scan performance because every record is stored in a linked list node, traversing them results in many cache misses. 

\textbf{Summary.}
The most effective technique to achieve high performance remains leveraging DRAM, 
which newer PM indexes adopt more aggressively, by putting more components or even complete trees in DRAM. 
This is at the cost of higher memory consumption, more complex recovery protocol and higher cost of ownership. 
Importantly, \fptree remains very competitive. 
\pactree and \roart are only marginally faster than \fptree for lookups. 
In Figure~\ref{fig:pm-uniform-1t}(b), \pactree, \roart and \utree are even slower than \fptree. 
Only \dptree and \pactree perform better than \fptree for scans.

\subsection{Multi-threaded Experiments}
\label{subsec:multi-thread-exp}
Now we measure index throughput with different thread counts under uniform and skewed distributions.  
We start with one socket and expand to NUMA with two sockets in Section~\ref{subsec:numa-exp}.

\textbf{Individual Operations.}
Figures~\ref{fig:pm-mt}(a--d) show the throughput of lookup/insert/update/scan under the uniform distribution; 
the shaded areas (over 20 threads) indicate numbers obtained when hyperthreads are also used. 
All indexes scale well for pure lookups in Figure~\ref{fig:pm-mt}(a), with \dptree and \lbtree achieving higher raw throughput than others. 
This result aligns with that obtained in Section~\ref{subsec:single-thread-exp} under a single thread. 
\dptree uses optimistic lock coupling for its buffer/base trees, and readers can traverse without incurring PM accesses. 
Its bloom filter also helps avoid unnecessary lookups in the buffer trees. 
\lbtree uses HTM which without write operations exhibits little/no aborts. 

For inserts, although \lbtree does not perform the best under a single thread, it scales the best under multiple threads, by being 1.55$\times$/1.96$\times$/3.07$\times$/2.23$\times$ faster than \roart-DCMM/\roart-PMDK/\dptree/\pactree.
Although \lbtree inherits many designs from \fptree, its logless split, new node layout and inner node locks to avoid re-traversals during split further make it 2.09$\times$ faster than \fptree.
\dptree's performance stops scaling beyond 10 threads, mainly due to its 7-phase merge: 
each time a merge occurs, records in the middle buffer tree will be moved into the base tree, causing the base tree's inner nodes to be rebuilt. 
This in turn incurs high garbage collection costs. 
\roart-DCMM achieves 1.27$\times$ higher performance using its customized PM allocator compared to using PMDK. 
Not leveraging DRAM also contributes to its lower performance compared to others that do leverage DRAM. 

For updates, \lbtree outperforms others in Figure~\ref{fig:pm-mt}(c), thanks to its fast traversal and node layout design. 
\dptree is also very competitive, as updates are served in-place without triggering merge operations. 
Similar to the single-threaded results, for all indexes, updates behave more similarly to inserts (than lookups) as leaf-level locks must be retained until the new value is flushed. 

As shown in Figure~\ref{fig:pm-mt}(d), under multiple threads the relative merits of different indexes on range scan are similar to the single-threaded results. 
The only exception and best performing index is \pactree. 
It scales to 40 threads, thanks to the combination of 
(1) its leaf node design that inlines key-value pairs and leverages PM's fast sequential read, 
(2) indirection that avoids sorting, 
and (3) optimistic concurrency that incurs no PM writes for reads. 
The other trie-based \roart performs the worst although it specifically optimizes for scan since it does not use DRAM and incurs more cache misses (Section~\ref{subsec:single-thread-exp}); 
under high core counts, cache misses are further exacerbated in Figure~\ref{fig:cache_misses_pmem}(h).
For B+-tree variants, \lbtree performs much worse for scans due to its use of small nodes (more cache misses). 
In contrast to the single-threaded results, \fptree performs poorly: 
using larger leaf reduces cache misses, but increases lock contention on leaf nodes. 
\dptree takes no locks for scans (OLC), but needs to visit multiple indexes and merge results, which contributes to its lower performance than \pactree. 

\begin{figure}[t]
	\centering
	\includegraphics[width=\linewidth]{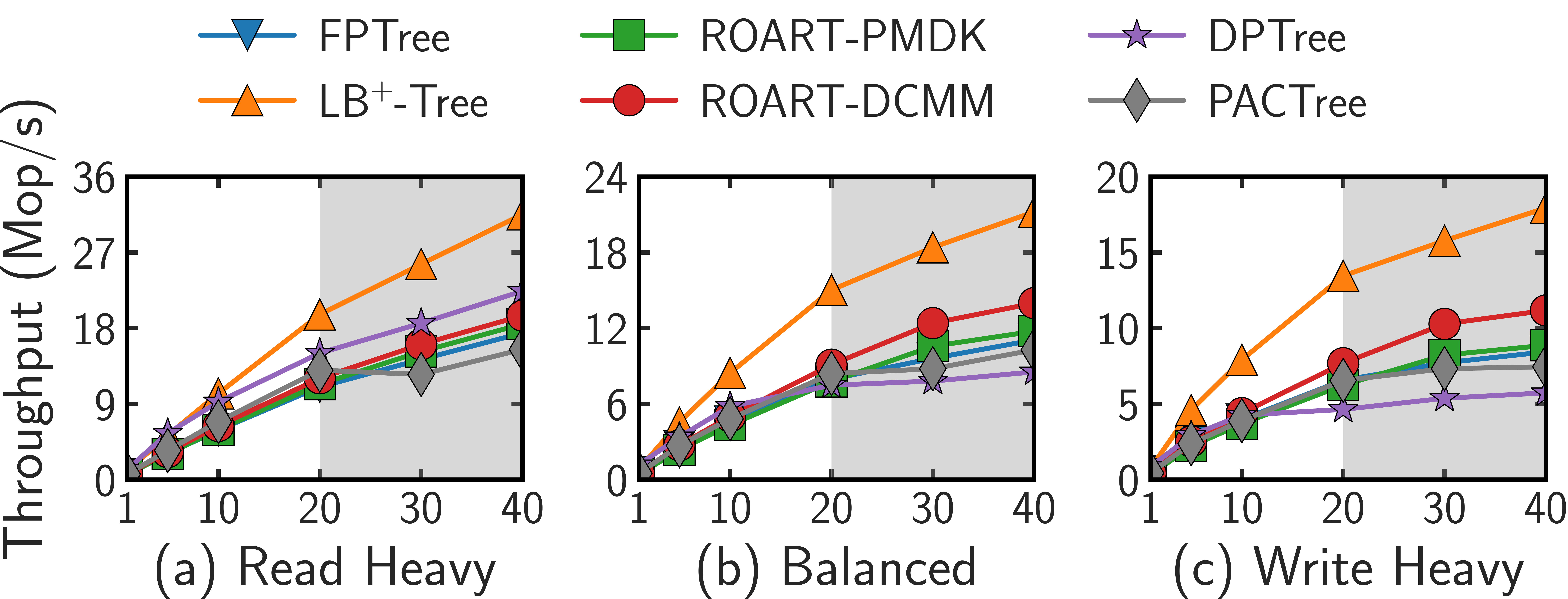}
	\caption{Throughput of mixed workloads (lookups + inserts) under uniform distribution.}
	\label{fig:mixed_workload_pmem}
\end{figure}

\begin{figure*}[t]
	\centering
	\includegraphics[width=\linewidth]{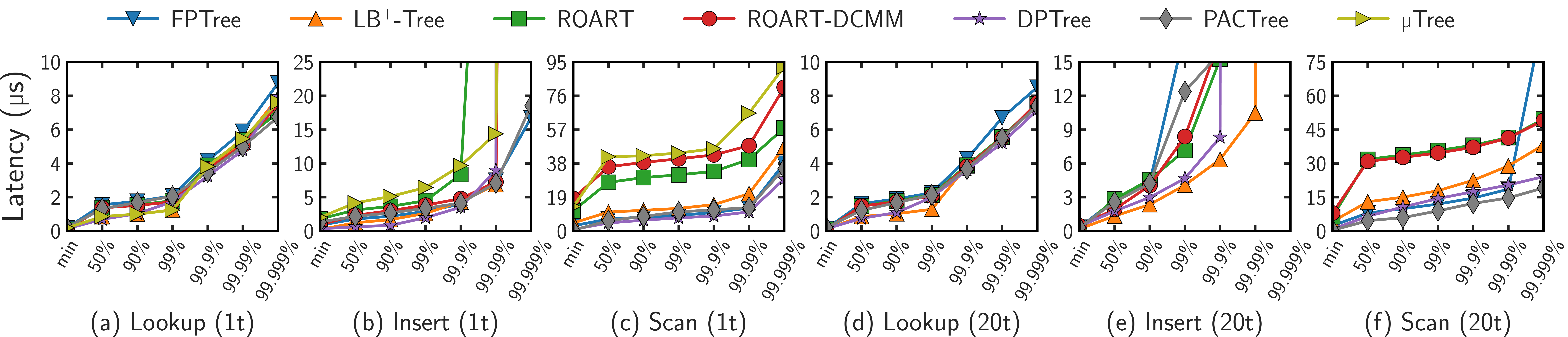}
	\caption{Tail latency of PM indexes under uniform distribution and a single thread (a--c) and 20 threads (d--f).}
	\label{fig:pm-latency}
\end{figure*}

\textbf{Skewed Accesses.}
Under the self-similar distribution where 80\% of accesses are focused on 20\% of all the keys~\cite{BillionSynthetic}, 
lookups as shown in Figure~\ref{fig:pm-mt}(e) exhibit a similar trend to uniform distribution but with higher raw throughput because of better CPU cache utilization (the working set is smaller). 
For updates in Figure~\ref{fig:pm-mt}(f), 
\dptree remains scalable as updates are all performed in the DRAM buffer tree.
\fptree shows unstable and low throughput for updates due to frequent HTM aborts and more PM writes caused by SMOs during update: updates to the same record are appended without deduplication, so the node can become full and get split during an update. 
No index scales under update workloads with hyperthreading. 
Under contention, locking takes over to become the main bottleneck in \fptree, despite the working set is smaller. 

Like \fptree, \lbtree also uses locking for leaf nodes, but scales under scan because of their different ways of using HTM: 
\fptree delegates HTM and locking to TBB (Section~\ref{subsec:pre-optane-designs}) which has a global fallback path after a pre-defined number (256) of aborts of HTM transactions, whereas \lbtree directly uses HTM instructions (e.g., \texttt{xbegin}/\texttt{xend}). 
Although scan is read-only, the first leaf node lock (in PM) is acquired inside the HTM transaction at the end of traversal. 
This incurs much contention under skewed workloads and triggers HTM aborts, leading \fptree to use the slow fallback path protected by a global mutex. 
\dptree's optimistic locks allow high scan throughput under skewed accesses as no locks are acquired. 
\pactree scales with the best performance under skewed accesses for scans, although the gain diminishes with more hyperthreads. 
\roart scales slightly worse in skewed update: for each update it creates a new leaf and replaces the original leaf pointer inside the leaf array, which becomes more expensive under contention.

\textbf{Mixed Workloads.}
We test mixed workloads with different read/insert ratios: read heavy (90\% lookups + 10\% inserts), balanced (50\% lookups + 50\% inserts) and write heavy (10\% lookups + 90\% inserts). 
As Figure~\ref{fig:mixed_workload_pmem} shows, \lbtree exhibits the best performance and scalability. 
\dptree scales worse with more inserts as tree merge becomes a major overhead. 
Finally, \fptree again remains very competitive with \roart, \pactree and \dptree.

\subsection{Tail Latency}
\label{subsec:latency}
We measure tail latency using the same approach from previous work to strike a balance between overhead and accuracy~\cite{PiBench}. 
In detail, we sample 10\% of all the operations during each run under uniform distribution to rule out the impact of CPU caches. 
Figure~\ref{fig:pm-latency} shows the tail latency at varying percentiles under one thread (a--c) and 20 threads (d--f). 
As expected, we observe no obvious differences between one and 20 threads for lookups. 

For inserts, lookups and scans in Figures~\ref{fig:pm-latency}(a--c), \utree shows consistently higher latency (and skyrockets at 99.99\% for inserts), although its key design goal is to reduce tail latency (Section~\ref{subsec:btree}). 
We observe the reason is in its use of PM-resident linked lists with per-record nodes.  
To handle an insert, \utree uses $\sim$2900 cycles to allocate a list node and $\sim$2200 cycles to complete a \CAS and cacheline flush to insert the allocated node into the linked list. 
In contrast, \lbtree only needs around 80 cycles to insert a key into a leaf node. 
Moreover, the use of linked lists in the leaf layer causes many cache misses during scans: 
as shown in Figure~\ref{fig:cache_misses_pmem}(d), \utree exhibits the highest cache miss ratio. 
\roart-DCMM exhibits lower latency than \roart-PMDK for inserts, thanks to the better DCMM allocator: in the worst case, a split in \roart could allocate 63 leaf arrays and one inner node.
\roart has relatively higher latency for scans in Figures~\ref{fig:pm-latency}(c) and \ref{fig:pm-latency}(f), due to pointer chasing at the leaf level to dereference pointers to keys. 
The other trie-based \pactree directly stores keys in leaf nodes, hence exhibiting lower latency. 
\pactree also shows relatively low latency for all operations in most cases (except inserts under 20 threads) because SMOs are offloaded to background threads. 
Under 20 threads, the background threads start to fall behind, requiring worker threads to traverse extra nodes to reach the correct leaf node, thus increasing latency.

\subsection{Support for Variable-Length Keys}
\label{subsec:varkey}

\begin{figure}[t]
  \centering
  \includegraphics[width=\linewidth]{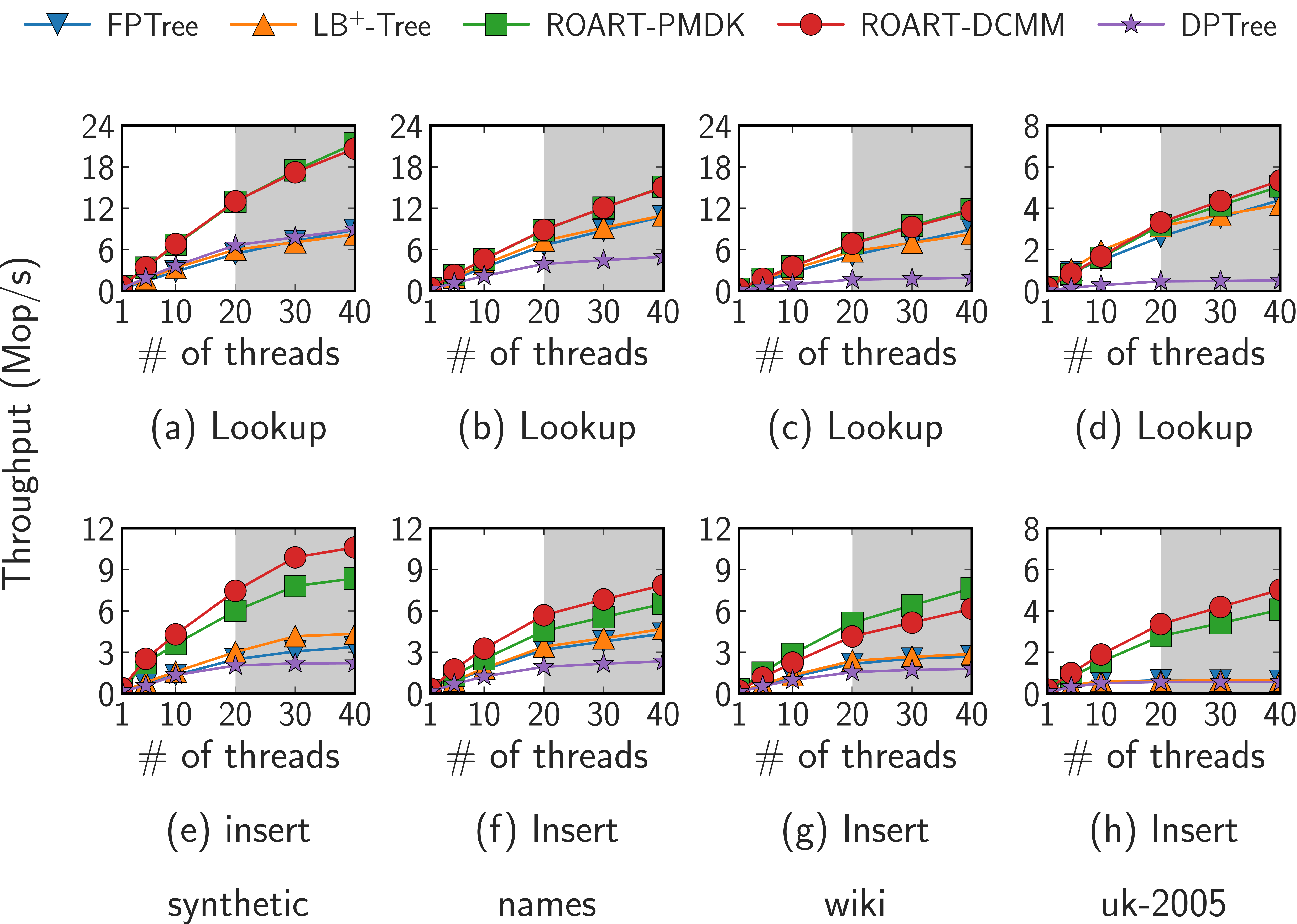}
  \caption{Throughput under variable-length keys using synthetic and real-world datasets~\cite{LastMile}. }
  \label{fig:var-key-pmem}
\end{figure}

Now evaluate variable-length key support.  
We first run the same experiment as Section~\ref{subsec:multi-thread-exp} with 8-byte keys, but force the indexes to use their variable-length key support. 
This may limit the depth of trie-based indexes (thus giving them advantages), but allows us to reason about the efficiency of the variable-length key support by comparing with the fixed-length key experiments; we use three real-world datasets later. 
For most indexes, including \fptree, \lbtree and \dptree this means allocating the key in the heap and storing a pointer to the key in the index. 
\roart uses its own native support to store keys in index nodes themselves. 
All indexes are plotted except \pactree.\footnote{\pactree assumes null-terminated strings. This is incompatible with \pibench which may generate keys with \texttt{$\backslash$0}, which will be wrongly treated as short keys.}
As shown in Figures~\ref{fig:var-key-pmem}(a) and~\ref{fig:var-key-pmem}(e), \roart performs the best in all cases. 
This is exactly opposite to the case using fixed-length keys (cf. Figure~\ref{fig:pm-mt}). 
In particular, cache misses caused by pointer chasing (to access keys) dominate the performance of \fptree, \lbtree and \dptree.  

We further test the indexes with three representative real-world datasets~\cite{LastMile}: Reddit usernames (\texttt{names})~\cite{dataset-reddit-usernames}, Wikipedia (\texttt{wiki})~\cite{dataset-wiki} and URLs (\texttt{uk-2005})~\cite{dataset-uk-urls}.
They respectively consist of string keys of up to 25, 256 and 2029 bytes; for space limitation we omit more details about the datasets which can be found in the Appendix A. 
As shown in Figures~\ref{fig:var-key-pmem}(b--d) and~\ref{fig:var-key-pmem}(f--h), the gaps between \roart and \fptree/\lbtree shrink, although \roart remains advantageous. 
With longer keys, trie-based indexes will build more inner nodes to form deeper traversal paths, while the depth of B+-tree variants is not affected as they store pointers to keys.
Among all the indexes, \dptree performs much worse with longer keys. 
We found a main reason is that 
most lookups (94\% according to our profiling results) needed to traverse the base tree and search leaf nodes. 
Since \dptree maintains multiple indexes, it also requires more complex traversal logic that incurs extensive key comparisons (on average 34 \texttt{memcmp} calls vs. 22 in \fptree), further lowering its performance. 

Overall, these results highlight the need to enhance variable-length key support in future PM indexes, for example by combining the best of tries and B+-trees without tradeoffs. 

\subsection{Impact of NUMA Effect}
\label{subsec:numa-exp}
Now we extend our experiments to use both NUMA nodes on the server. 
Except for \pactree, we allocate PM and DRAM from the first socket. 
All the threads are pinned, and we first use all the 40 physical cores across two sockets, before using hyperthreading beyond 40 threads. 
This allows us to stress the indexes with inter-socket traffic and contrast their behavior with and without NUMA effect. 
For \pactree, we include two variants which respectively enable and disable its NUMA-aware per-node PM pools. 

\begin{figure}[t]
	\centering
	\includegraphics[width=\linewidth]{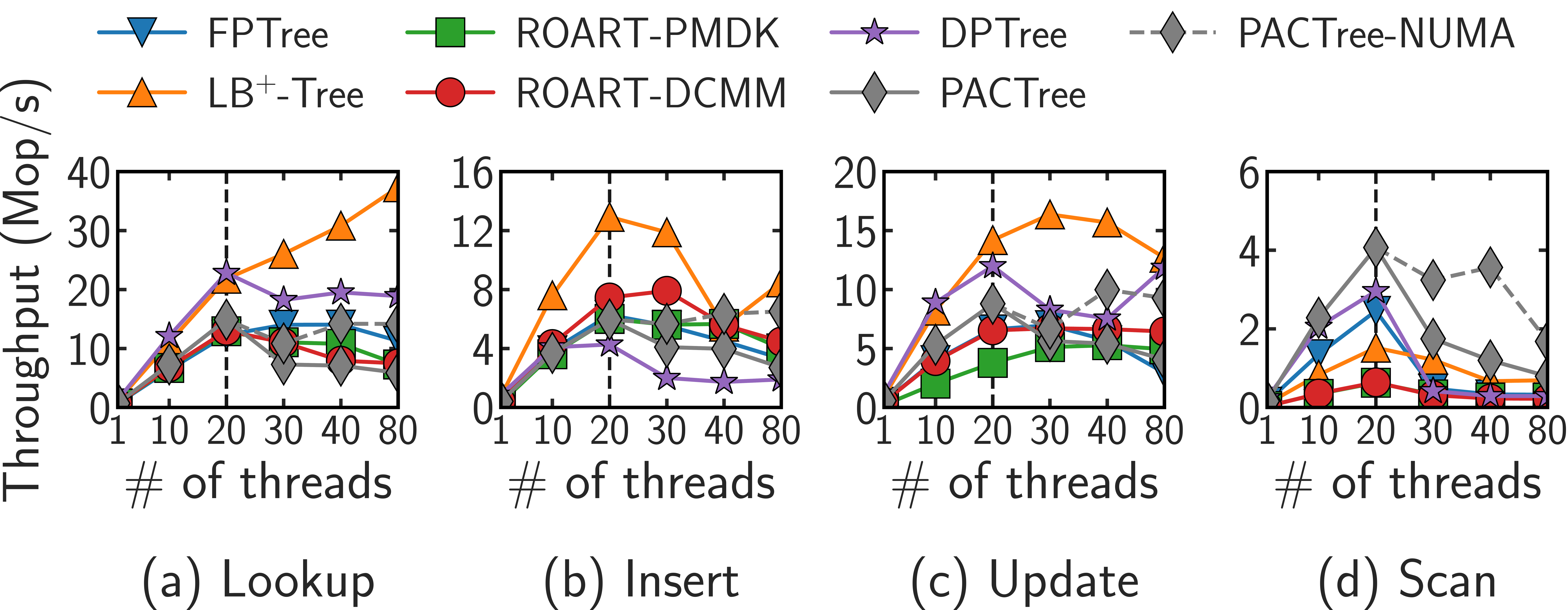}
	\caption{Impact of NUMA effect for PM indexes. 
  No index scales well under all operations due to additional PM accesses by the directory-based CPU coherence protocol. 
  }
	\label{fig:numa-pm}
\end{figure}

As shown in Figure~\ref{fig:numa-pm}, NUMA effect has major impact on all indexes' throughput, and no index scales well beyond one socket for all operations. 
Although not specifically designed for NUMA, \lbtree achieves the best scalability for lookups, and the highest throughput for inserts and updates (with a dropping trend beyond one socket). 
We attribute the reason to its frugal use of cacheline flushes and careful node layout designs. 
Both reduce PM accesses and cross-socket traffic. 
For lookups, \fptree also does not collapse, whereas the performance of other indexes fluctuate and/or drop beyond 20 threads. 
This implies HTM is robust to NUMA effect for read-only workloads, thanks to its lightweight conflict detection mechanism that piggybacks on the coherence protocol. 
Moreover, HTM can use the extra last-level cache in the second NUMA node to track reads~\cite{ParkHTM2018,UnderstandingHTMCapacity}. 
Other approaches (OLC, locking, ROWEX) are unable to take good advantage of the coherence protocol like HTM, contributing to more severe NUMA effect.

\pactree is the only PM index that takes NUMA effect into account by (1) using separate PM pools for each NUMA node and (2) leveraging snooping coherence protocol. 
With separate PM pools (\texttt{\pactree-NUMA}), \pactree maintains performance beyond one socket without collapsing, but still does not scale as expected.
The main culprit is the directory-based coherence protocol that incurs additional PM accesses to update the directory. 
Thus, \pactree advocates using snooping protocols for PM, which broadcasts coherence traffic across all cores, instead of using a directory to record cacheline status, hence does not incur additional PM accesses.
However, most platforms default to a directory-based protocol because snooping may not scale to high core counts. 
Since our server 
does not allow changing coherence protocols, we were unable to verify the performance of PM indexes using snooping protocols; we leave it as future work. 
As we have noted in Section~\ref{subsec:trie}, requiring a certain coherence protocol may inflict issues with other applications and limit the applicability of the index. 

\subsection{PM and DRAM Space Consumption}
New PM indexes are using DRAM more extensively to achieve high performance. 
The downsides are (1) more complex recovery protocols and (2) higher DRAM space consumption (hence higher cost of ownership). 
Table~\ref{tbl:mem-usage} lists the amount of DRAM and PM used by each index after loading 100 million records of 8-byte key and values (1.6GB). 
\dptree and \utree practically store complete trees in DRAM, resulting in up to $\sim$18$\times$ higher DRAM consumption when compared with \fptree and \lbtree.
\pactree uses a similar amount of PM to \utree's and is among the most frugal in using DRAM (second to \fptree) because of its packed design. 

Surprisingly, \roart uses 18.92GB to index 1.6GB of data, while others need 2.5--4.8GB. 
The reason is it requires cacheline-aligned nodes and overprovisions leaf arrays: 
for 100 million records, it allocates space for around one billion records (17 million leaf arrays) occupying 10.89GB of PM as each leaf array is 640-byte. 
However, most space (reserved for leaf pointers) are unused. 
Such overprovisioning is due to \roart's split mechanism. 
In our experiment, keys are uniform randomly generated, so the 64 records in a full leaf array are usually very different despite they share a common prefix. 
Then a split operation could allocate 63 leaf arrays in the worst case, leaving each new leaf array with only few records, leading to an average occupancy of $\sim$9\% and a waste of over 9GB of PM.  

\begin{table}[t]
\smallskip\noindent
\setlength\tabcolsep{0.6mm}
\caption{PM and DRAM consumption (GB) after loading 100 million records with 8-byte keys and 8-byte values.}
\label{tbl:mem-usage}
\begin{tabular}{lccccccc} 
\toprule
& \bf\fptree & \bf\lbtree & \bf\roart &  \bf\dptree & \bf\pactree & \bf\utree \\\midrule
{\bf DRAM} &  0.14  &  0.34  &  0.14 	&	1.14 & 0.18 & 2.63 &\\
{\bf PM} &  2.69  &  2.54  &  18.92    &	4.79 & 3.2 & 3.2     &\\
\bottomrule
\end{tabular}
\end{table}

\subsection{Bandwidth Utilization and Requirements}
\begin{figure}[t]
	\centering
	\includegraphics[width=\linewidth]{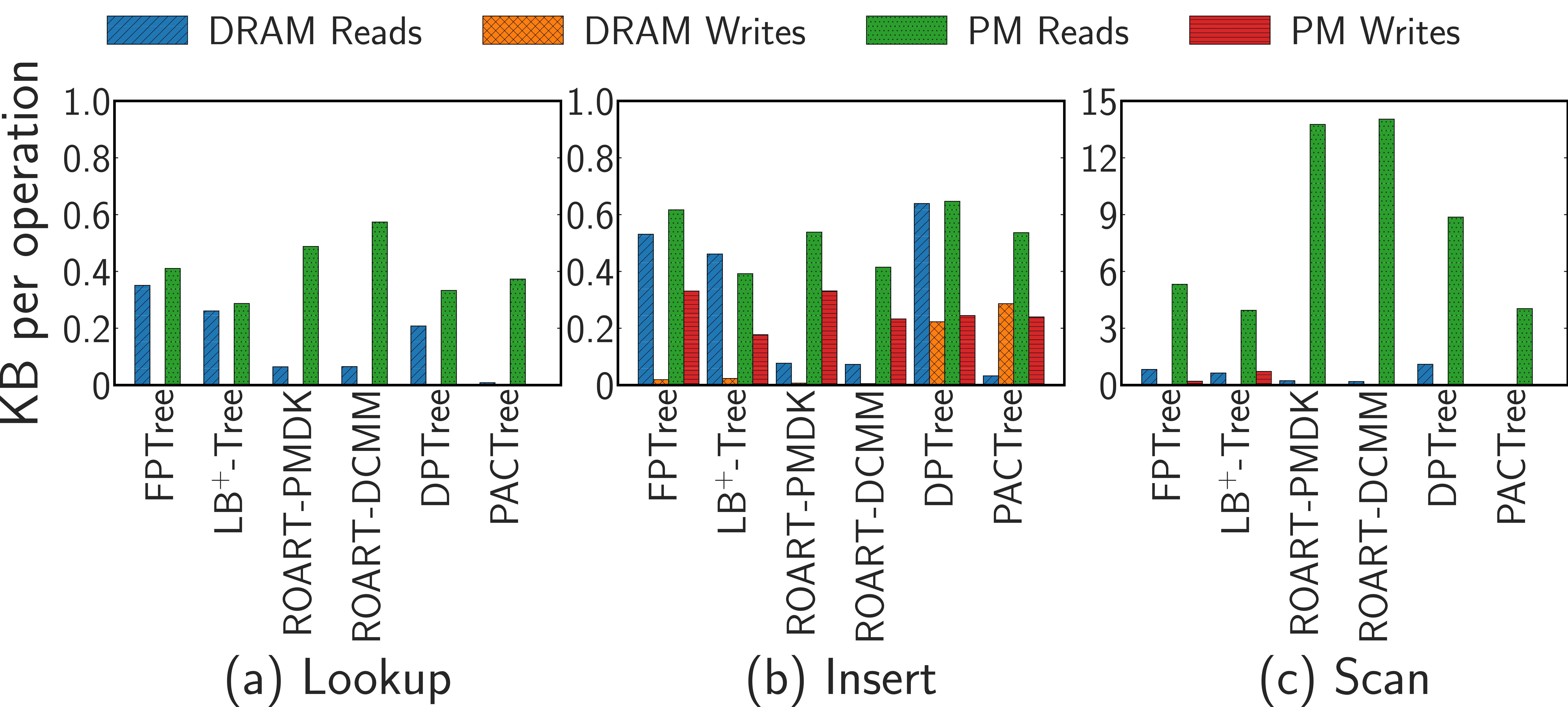}
	\caption{Bandwidth consumption per operation under 20 threads with 100 million records of 8-byte keys and values.} 
	\label{fig:pmbw}
\end{figure}

Previous evaluation~\cite{PiBench} has shown that PM bandwidth is scarce. 
Since a database system also uses various other components, it is desirable to keep the bandwidth consumption low for indexes. 
This has been the main focus of newer indexes. 
We observe that the peak usage across all indexes and operations does not reach the limit ($\sim$10GB/s/$\sim$40GB/s for random write/sequential read). 
This shows the effectiveness of the bandwidth saving techniques proposed by the surveyed indexes. 
Due to space limitation, we omit the details on total bandwidth utilization and focus on the bandwidth used per operation, which is more indicative on how frugal (or not) an index uses PM bandwidth. 
Figure~\ref{fig:pmbw} shows the results 
with 20 threads when running lookups, inserts and scans under 8-byte keys and 8-byte values under the uniform distribution. 
Overall, \lbtree exhibits the lowest number of bytes per operation, thanks to its node layout and logless design.  
\roart exhibits the highest PM reads as it overprovisions node space.
In contrast, \pactree, which is also trie-based, has a similar bandwidth requirement to \lbtree's, because of its layout designed to reduce unnecessary PM accesses.

\section{Observations and Insights}
\label{sec:observe}
In this section, we summarize the major findings based on our experimental results and analysis of the surveyed PM indexes. 

\textbf{1. The rule of thumb remains reducing PM accesses, which was first set in the pre-Optane era.} 
Almost all the design choices (both pre-Optane and new ones) center around this goal, due to PM's lower bandwidth and higher latency. 
If the properties of future PM hardware changes, the principles may be revisited. 

\textbf{2. Some building blocks from the pre-Optane era continue to work well and are further optimized by new PM indexes.} 
Most indexes use DRAM to accelerate traversal; some (e.g., \dptree) even place entire trees in DRAM. 
However, tries typically cannot take the full advantage of DRAM to store reconstructible data. 
Fingerprints are also widely used and enhanced by placing them in the spare bits of pointers and accessing them using SIMD instructions. 

\textbf{3. Using extra metadata and selective persistence of metadata can further accelerate performance.} 
The main reason is these approaches can avoid using WAL, which may incur additional PM writes and complicate code logic.

\textbf{4. Newly Proposed $\ne$ Better.} 
Pre-Optane \fptree is still very competitive and sometimes can even outperform newer indexes. 
Although \utree optimizes for tail latency, it exhibits the highest latency in many cases.  
Such results call for careful benchmarking and comprehensive evaluations. 

\textbf{5. All the new PM indexes are tailor-made for one product (Intel Optane PMem), which can be a double-edged sword.} 
While this can deliver high performance, as exemplified by \lbtree which performs the best in most cases, it could pose challenges when the PM hardware landscape becomes more diverse. 

\textbf{6. Support for NUMA-awareness, efficient PM management and variable-length keys remains inadequate.} 
There have been initial attempts (e.g., using pointers for variable-length keys), but they are usually ad hoc or partial solutions with practical limitations (e.g., requiring a specific coherence protocol). 

\textbf{7. There is no clear ``winner'' index architecture, but the choice may affect how (efficiently) functionality can be supported.} 
For example, B+-tree (trie) variants perform well for fixed-length (variable-length) keys. 
But 
it remains to be explored whether it is easier to add efficient variable-length key support in \lbtree or to optimize \pactree to match \lbtree's performance. 

\textbf{8. Linked lists with small nodes are a bad fit for PM indexes, and cache misses in general should be minimized or hidden.} 
Accessing and scanning through a linked list of individual records incur many cache misses which can dominate the performance and lead to high latency (e.g., in \utree), canceling out the positive effects brought by other optimizations. 

\textbf{9. HTM can perform well under NUMA for read-only workloads, but is challenging to handle contention and debug.} 
In particular, the programmability issue 
is further complicated with other system-level infrastructure: we observed extremely high abort rates under \texttt{glibc} version 2.33 which does not use the right instructions that can work with HTM in \texttt{memcpy}.\footnote{Details at \url{https://sourceware.org/bugzilla/show_bug.cgi?id=28033}.}
The bug was fixed in \texttt{glibc} 2.34~\cite{glibc234} which is used in our experiments. 
It is noticeable that the best performing \lbtree is based on HTM; 
it therefore remains to be seen in future work whether other approaches can overcome these issues while maintaining high performance.

\begin{figure*}[t]
	\centering
	\includegraphics[width=\linewidth]{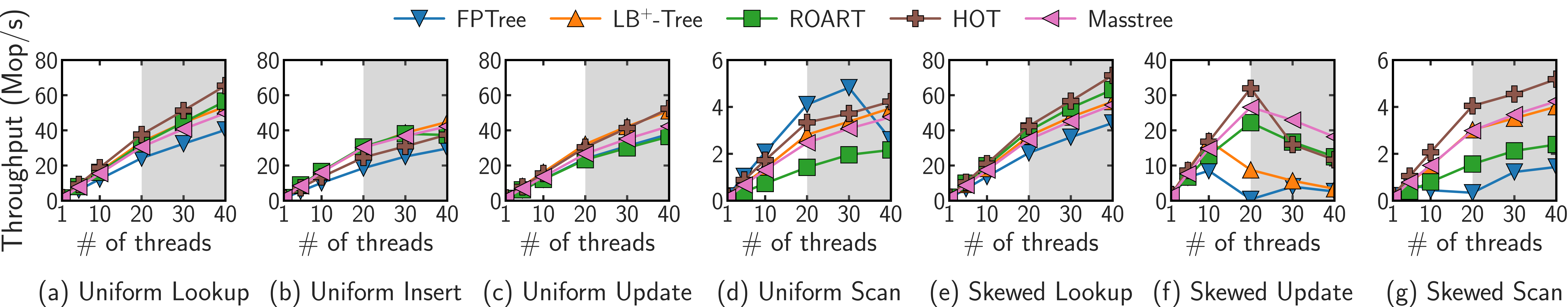}
	\caption{Throughput of PM and DRAM indexes running on DRAM. 
  Without the extra fences, cacheline flushes and PM management code, PM-tailored indexes at least match the performance of \hot and \masstree. 
  \lbtree performs even better than \hot and \masstree on certain insert workloads (b), and \fptree tops scan performance under uniform distribution (d). 
  }
	\label{fig:dram-tp}
\end{figure*}

\section{On the Next PM and DRAM Indexes}
\label{sec:next}
We give the outlook of the PM indexing space and describe promising future directions for PM and DRAM indexing. 

\subsection{Future PM Indexes}
We identify three promising areas of future work for PM indexes. 

\textbf{1. Efficient Support for Full Functionality.}
As we have discussed previously, variable-length keys and NUMA-awareness remain open problems for future PM indexes. 
Importantly, it is desirable to maintain the high performance obtained by existing designs while better supporting full functionality. 

\textbf{2. Wider Applicability/Less Tailor-Made.} 
There are various ways to realize PM, by using new materials (e.g., memristor~\cite{Memristor}, STT-RAM~\cite{STT-RAM} and Intel 3D XPoint which PMem is based upon) or NVDIMMs which combine flash and DRAM~\cite{AgigaNVDIMM,VikingNVDIMM}. 
However, most (if not all) indexes aiming for real PM are tailor-made for Intel Optane PMem; 
yet certain properties like 256-byte alignment may even change across generations of the same product, and designs based on them may not work well on NVDIMMs, diminishing their applicability. 
Although some of the hardware efforts are in their early stage, we argue it is important to consider applicability of future designs on different PM devices. 

\textbf{3. Real-World Adoption and Cost-Effectiveness.}
Although there have been numerous PM index proposals, we are yet to see major adoption in real systems and applications. 
Part of the reason is the low cost effectiveness of PM-based servers as identified by other work~\cite{SSDStrikingBack}, especially when compared to modern SSDs which can deliver high bandwidth and microsecond-level latency. 
Therefore, on the hardware side, we hope future work to lower the per GB cost of PM servers. 
On the software side, PM indexes and data structures in general should focus more on cost/performance. 

\subsection{Unifying PM and DRAM Indexing} 
In a similar vein to the point on wider applicability, we observe techniques proposed for PM indexes can also be effective for DRAM. 
We conduct preliminary experiments to compare the surveyed state-of-the-art PM indexes (with the extra cacheline flushes and fences removed) and two representative DRAM-optimized volatile indexes (\hot~\cite{HOT} and \masstree~\cite{Masstree}).
Figure~\ref{fig:dram-tp} shows the throughput obtained by running the same workload as Section~\ref{subsec:multi-thread-exp}, but purely on DRAM. 
Under both uniform and skewed distributions, PM indexes perform competitively with DRAM-optimized indexes. 
In certain cases, PM indexes perform even better than \hot and \masstree which are specifically optimized for DRAM, e.g., 
\lbtree for inserts in Figure~\ref{fig:dram-tp}(b) and \fptree (despite being a pre-Optane proposal) in Figure~\ref{fig:dram-tp}(d) for scans. 
Although it remains to be seen how the optimizations for PM and DRAM indexes compare, and how PM techniques may be used by DRAM indexes (and vice versa), our results indicate it is promising to devise future indexes that would work on both volatile and non-volatile memory. 
This could greatly simplify implementation and widen the applicability of techniques proposed by both types of indexes.

\section{Related Work}
\label{sec:related}
Our work is most related to performance studies for PM devices and data structures, PM indexes and PM management issues.

\textbf{Performance Studies.} 
Early work~\cite{UCSDMeasurement,UCSDGuide} characterized the performance of Optane PMem, exposing a set of properties different from what were previously assumed by emulations. 
Gugnani et. al~\cite{PMIdio} exposed more properties of Optane PMem under various scenarios, e.g., eADR and NUMA, along with case studies. 
Beyond range indexes, Hu et. al~\cite{HuPMHI} evaluated PM-optimized hash indexes on Optane PMem. 
Koutsoukos et. al~\cite{HtuPMiD} analyzed the performance of PM-enhanced database engines under TPC-C and TPC-H, and came up with guidelines of tuning the system for best performance. 

\textbf{PM Indexes.}
In addition to adapting specific indexes, general-purpose approaches such as 
RECIPE~\cite{RECIPE}, NAP~\cite{NAP} and TIPS~\cite{TIPS} present principled methods for converting DRAM indexes into PM indexes. 
It is interesting future work to evaluate these approaches.  
Some early efforts have adapted learned indexes~\cite{LearnedIndex} for PM. 
Chen et. al~\cite{PMLearnedEval} observe that the bigger nodes used by learned indexes can cause excessive PM accesses. 
APEX~\cite{APEX} transforms the DRAM-based updatable ALEX~\cite{ALEX} with concurrency and instant recovery on PM. 
Hash tables are also being re-designed for PM. 
CCEH~\cite{CCEH} is a failure-atomic variant of extendible hashing that reduces directory management overhead. 
Dash~\cite{Dash} integrates a set of useful techniques to adapt extendible and linear hashing for PM; 
the key insight is that both PM reads and writes should be minimized. 
Clevel~\cite{CLEVEL} is a lock-free version of level hashing~\cite{LevelHashing} that performs asynchronous resizing in the background. 

\textbf{PM Libraries.} 
PMDK~\cite{PMDK} is the de facto standard, but may not be the optimal solution:  
\roart, \dptree and \pactree all devise their own approaches. 
Designing better PM libraries remains an open area, as seen by many recent alternatives~\cite{Pangolin,Ralloc,Metall,Makalu,nvmMalloc}.

\section{Conclusion}
\label{sec:conclusion}
We conducted a comprehensive evaluation of representative PM range indexes proposed based on real Intel Optane PMem. 
These new indexes inherited many useful designs from pre-Optane PM indexes and proposed new building blocks 
that can be useful for building future PM indexes. 
Based on our evaluation, we gave a list of observations, insights and future directions. 
We found the new indexes do not necessarily outperform the pre-Optane proposals, and efficient designs for variable-length keys, PM management and NUMA awareness are still lacking. 
We also discovered for the first time that a PM range index can match or even outperform DRAM-optimized indexes, highlighting the potential of unifying PM and DRAM indexing to save design and implementation efforts.

\begin{acks}
We thank the anonymous reviewers and associate editor for their constructive feedback. 
This work is partially supported by an NSERC Discovery Grant, Canada Foundation for Innovation John R. Evans Leaders Fund and the B.C. Knowledge Development Fund.
\end{acks}

\bibliographystyle{ACM-Reference-Format}
\bibliography{ref}

\begin{appendix}

\begin{figure*}[t]
	\centering
	\includegraphics[width=0.33\linewidth]{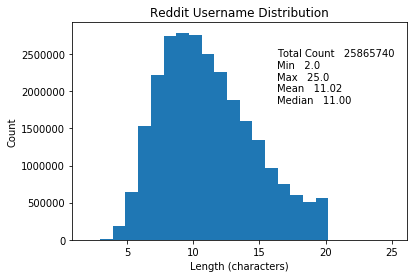}
	\includegraphics[width=0.33\linewidth]{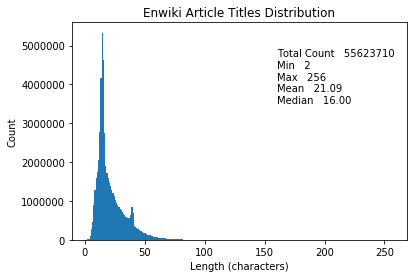}
	\includegraphics[width=0.33\linewidth]{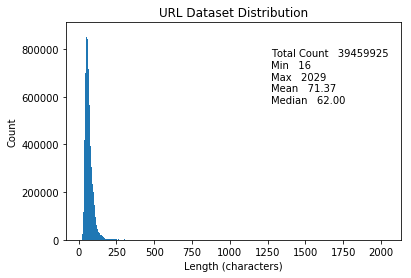}
  \caption{Key length distributions of three real-world variable-length key datasets.}
  \label{fig:dist}
\end{figure*}
\section*{Appendix}

\section{Datasets Used and Clarification}
Here we present the details of all three datasets used in the variable-length key experiments. 
The \texttt{names} dataset contains 25 million unique Reddit usernames of length up to 25
characters. 
\texttt{wiki} contains 55 million unique article titles of length up to 256 characters.
\texttt{uk-2005} contains 40 million unique URLs of up to 2029 characters. 
The key distributions and statistics of the three datasets are shown in Figure~\ref{fig:dist}, covering a wide range of different key lengths in reality. 
As a trie-based index, ROART already supports variable length keys.
So no modification is required. 
For \lbtree, \fptree and \dptree, we modified their lookup and insert operations to support variable-length keys. 
The parameters of the related operations are unchanged but we allocate the variable-length keys from volatile/persistent heap and store the key pointers in inner/leaf nodes. 
Key comparison is performed using \texttt{memcmp} over the two key pointers. 
We also allocate each key by the maximum length regardless of its actual length, to ease implementation. 
This approach does results in wasted memory especially in the case of skewed datasets.
 \begin{figure}[t]
	\centering
	\includegraphics[width=\linewidth]{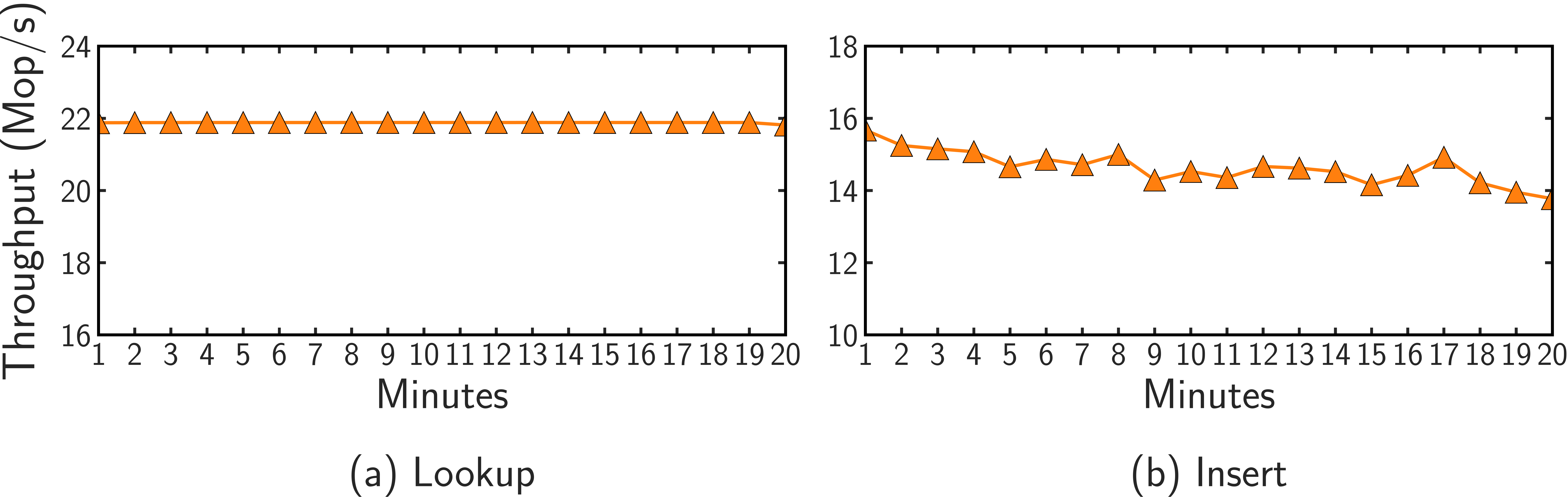}
	\caption{\lbtree performance over time.}
  \label{fig:time-perf}
\end{figure}

\section{Long-Running Workloads} 
Although PM's bandwidth exhausts quickly with increasing number of writers~\cite{PiBench}, its performance tends to be stable across the short-lived benchmarks. 
To simulate the impact of long-running systems and high space utilization on PM performance, we conducted two stress tests using \lbtree. 

A pool of 485GB is first allocated from PM, which is the maximum available PM space that can be allocated on the first node of our machine.  
We then populated the entire reserved space with 8-byte keys/values generated from a uniform distribution, and ran a read-only workload for 20 minutes with 40 threads. 
The lookup throughput averaged by minutes is shown in Figure~\ref{fig:time-perf}(a). 
As depicted, the number of per-second read operations completed is very stable across the entire experiment phase. 

We further conducted another insert experiment, where 20 billion uniformly distributed 8-byte key/values are inserted with 40 threads. 
The experiment kept running until the PM space is exhausted (it took around 20 minutes). 
As shown by Figure~\ref{fig:time-perf}(b), the insert throughput was also stable with minor gradual degradation ($\sim$12\%) as we insert more keys. 
This is expected because the index gets larger and deeper with more keys inserted, thus resulting in slower traversal and SMOs. 

Overall, PMem demonstrates robustness and stable performance throughout the long-running read/write intensive benchmarks while under high space utilization. 
Different from traditional SSDs that are known to perform worse as approaching full space utilization, PMem has shown no obvious hardware/software performance degradation (except due to the characteristics of the workload itself, e.g., inserts) in our experiments. 
However, we acknowledge that these tests are preliminary and further investigation with even longer experiments would be needed to thoroughly explore this topic in future work. 

\section{NUMA Experiments on DRAM}
\begin{figure}[t]
	\centering
	\includegraphics[width=\linewidth]{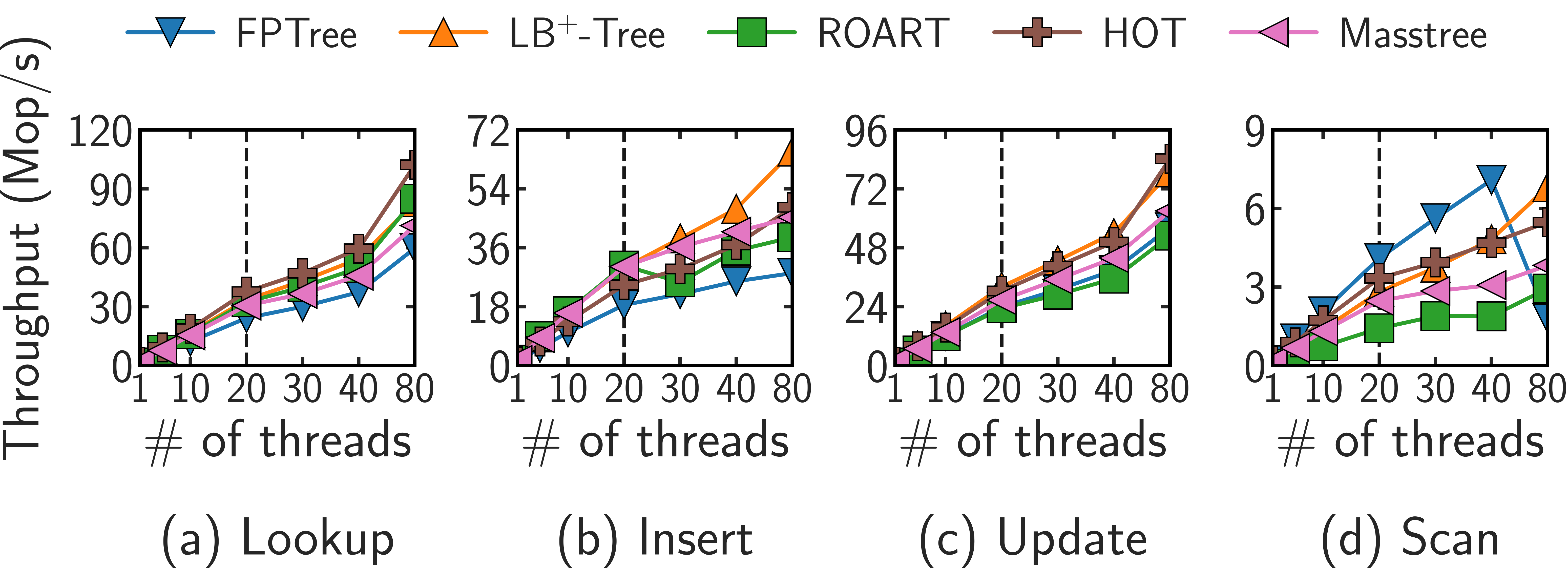}
	\caption{Impact of NUMA effect for indexes on DRAM.}
	\label{fig:dram-numa}
\end{figure}

This set of experiments repeats our earlier NUMA experiments in the main text but were done using DRAM to explore the impact of NUMA effect. 
As shown by Figure~\ref{fig:numa-pm}, most PM indexes are not scalable and their throughput can drop drastically beyond one socket.
However, it is not the case on DRAM. 
As Figure~\ref{fig:dram-numa} shows, when running on DRAM, \fptree, \lbtree and \roart are still scalable across two NUMA nodes. 
Thus, the main PM-specific concern lies in the impact of PM's higher latency/lower bandwidth on the interconnect traffic and cache coherence protocol~\cite{pactree}. 

\end{appendix}

\end{document}